\documentclass[11pt,a4paper,nofootinbib,showpacs,preprint,utf8]{article}
\pdfoutput=1 

\usepackage{jheppub}
\usepackage[english]{babel}
\usepackage{amsmath,amssymb,amsfonts, bm,bbm,slashed, subdepth}
\usepackage{graphicx}
\usepackage{enumerate}
\usepackage{setspace}
\usepackage{booktabs, tabularx}
\usepackage{units}
\usepackage{color}
\usepackage{float}
\usepackage{multirow}
\usepackage{array}
\usepackage{makecell}
\newcolumntype{C}[1]{>{\centering\arraybackslash}m{#1}}
\usepackage[dvipsnames]{xcolor}
\usepackage[pscoord]{eso-pic}
\usepackage[normalem]{ulem}
\usepackage{import}
\usepackage{url}
\usepackage{subcaption}
\usepackage{tikz}
\usetikzlibrary{snakes}
\usepackage{cancel}
\allowdisplaybreaks[1]

\usepackage{scalerel}
\usepackage{soul}
\usepackage{comment}

\usepackage{hyperref}
\hypersetup{
	setpagesize=false,
	bookmarksnumbered=true,
	colorlinks=true,
	linkcolor=blue,
	citecolor=red,
	hypertexnames=true
}

\definecolor{lime}{HTML}{A6CE39}
\DeclareRobustCommand{\orcidicon}{\hspace{-2.1mm}
	\begin{tikzpicture}
	\draw[lime,fill=lime] (0,0.0) circle [radius=0.13] node[white] {{\fontfamily{qag}\selectfont \tiny ID}}; \draw[white,fill=white] (-0.0525,0.095) circle [radius=0.007]; 
	\end{tikzpicture} \hspace{-3.7mm} }
\foreach \x in {A, ..., Z}
{\expandafter\xdef\csname orcid\x\endcsname{\noexpand\href{https://orcid.org/\csname orcidauthor\x\endcsname} {\noexpand\orcidicon}}}

\newcommand{\be}{\begin{equation}}
	\newcommand{\ee}{\end{equation}}
\newcommand{\ba}{\begin{array}}
	\newcommand{\ea}{\end{array}}
\newcommand{\bea}{\begin{eqnarray}}
	\newcommand{\eea}{\end{eqnarray}}
\newcommand{\balg}{\begin{align}}
	\newcommand{\ealg}{\end{align}}
\newcommand{\bit}{\begin{itemize}}
	\newcommand{\eit}{\end{itemize}}

\definecolor{bostonuniversityred}{rgb}{0.8, 0.0, 0.0}

\makeindex

\title{Singlet-Doublet Fermionic Dark Matter in Gauge Theory of Baryons}
 \author[a]{Taramati}
 \author[b,c]{, Rameswar Sahu\orcidA{}}
 \author[a]{, Utkarsh Patel}
\author[b,c]{, Kirtiman Ghosh}
\author[a]{, Sudhanwa Patra}
\affiliation[a]{Department of Physics, Indian Institute of Technology Bhilai, Durg 491002, India}
\affiliation[b]{Institute of Physics, Bhubaneswar, Sachivalaya Marg, Sainik School, Bhubaneswar 751005, India}
\affiliation[c]{Homi Bhabha National Institute, Training School Complex, Anushakti Nagar, Mumbai 400094, India}
\emailAdd{taramati@iitbhilai.ac.in}
\emailAdd{rameswar.s@iopb.res.in}
\emailAdd{utkarshp@iitbhilai.ac.in}
\emailAdd{kirtiman.ghosh@iopb.res.in}
\emailAdd{sudhanwa@iitbhilai.ac.in}
\begin{document}

\abstract{We are considering a minimal $U(1)_B$ extension of the
Standard Model (SM) by promoting the baryon number as a local gauge
symmetry to accommodate a stable dark matter (DM) candidate. The gauge theory of baryons induces non-trivial triangle gauge anomalies, and we provide a simple anomaly-free solution by adding three exotic fermions. A scalar $S$ spontaneously breaks the $U(1)_B$ symmetry, leaving behind a discrete $Z_2$ symmetry that ensures the stability of the lightest exotic fermion was originally introduced to cancel the triangle gauge anomalies. Scenarios with weakly interacting DM candidates having non-zero hypercharge usually face stringent constraints from experimental bounds on the DM spin-independent direct-detection (SIDD) cross-section. In this work, we consider a two-component singlet-doublet fermionic dark matter scenario, which significantly relaxes the constraints from bounds on the DM SIDD cross-section for suppressed singlet-doublet mixing. We show that the model offers a viable parameter space for a cosmologically consistent DM candidate that can be probed through direct and indirect searches, collider experiments, and
gravitational wave (GW) experiments.}

	\maketitle
\newpage
\section{Introduction}
\label{sec:intro}
The standard Model~(SM)~\cite{Glashow:1961tr,Weinberg:1967tq,Salam:1968rm,ParticleDataGroup:2012pjm} of particle physics has been highly successful in describing the interactions of the fundamental constituents of matter, and its predictions have been tested to a very high degree of accuracy in various experiments~\cite{CDF:1995wbb,ATLAS:2012yve,CMS:2012qbp}. Particle interactions in the SM are described within the framework of quantum mechanics and the theory of relativity, more precisely by a local relativistic quantum field theory. Each fundamental particle, treated as pointlike, is associated with a field with suitable transformation properties under the Lorentz group. The description of all the particle interactions in the SM is based on a common principle: \textbf{``Gauge"} invariance~\cite{Weyl:1929fm,Yang:1954ek}. Gauge invariance suggests that the lagrangian describing the system remains invariant under the local transformation of certain symmetry groups.
\par Despite its success at many frontiers, the SM does not explain certain phenomena such as gravitational interaction of matter~\cite{Einstein:1916vd}, matter-anti matter asymmetry~\cite{Planck:2018vyg}, non-zero neutrino mass and mixings~\cite{Super-Kamiokande:1998kpq,SNO:2002tuh,K2K:2002icj}, the existence of dark matter~\cite{Julian:1967zz,SDSS:2003eyi}, etc. This paves the way for the theories that fall under the category of Beyond Standard Model~(BSM) Physics. Given the gauge-invariant nature of SM, it seems highly logical to build BSM theories where the SM gauge symmetry is extended at higher energies with the usual gauge structure retained at the low-energy scales. This motivates researchers to put forward various BSM theories where the SM at higher energy scales is extended with additional unitary gauge groups~\cite{Pati:1973uk,Pati:1974yy,Mohapatra:1974gc,Mohapatra:1974hk,Georgi:1974sy,Georgi:1974yf,Georgi:1974my,DeRujula:1975qlm,Fritzsch:1974nn,Pais:1973mi,Duerr:2013dza,Patel:2022qvv}.
\par Baryon and lepton number of fundamental particles are accidental global symmetries of the SM. This alludes to their promotion as local gauge symmetries at higher energy scales. This was first introduced in Ref.~\cite{FileviezPerez:2010gw} and followed upon in works of Ref.~\cite{FileviezPerez:2011pt,Dulaney:2010dj}. Such attempts have also been made in literature in connection with the quantum theory of gravity~\cite{Kallosh:1995hi,Banks:2010zn}. Thus, following on the similar line of works, here we consider a $U(1)_B$ extension of the SM, where the baryon number~$(B)$ is promoted as a local gauge symmetry. The details of our model framework can be referred to in section~\ref{sec:gaugeB}.
\par Dark Matter~(DM) has been a mystery ever since its inception~\cite{Zwicky:1933gu,Rubin:1980zd}. It, along with other SM limitations, motivates a plethora of BSM frameworks. With the introduction of the WIMP~(Weakly Interacting Massive Particle) scenario~\cite{Kolb:1990vq}, the search for DM candidates in the GeV$\sim$TeV mass range became prominent. Despite the recent advancement in DM direct-detection experiments such as LUX~\cite{daSilva:2017swg}, XENON-100~\cite{XENON100:2012itz}, PANDAX-II~\cite{PandaX-II:2016vec,PandaX-II:2017hlx}, XENON-1T \cite{XENON:2015gkh, XENON:2018voc} and indirect-detection experiments such as PAMELA~\cite{PAMELA:2013vxg,PAMELA:2011bbe}, AMS-2~\cite{Corti:2014ria}, Fermi Gamma-ray space Telescope~\cite{Fermi-LAT:2009ihh} and IceCube ~\cite{IceCube:2017rdn,IceCube:2018tkk} to name few, there has been no hint of any excess of signal events over the background indicating a DM particle signature. This has led to a highly constraint parameter space for BSM scenarios with a DM candidate detectable at these experiments.

Our work builds on techniques developed to circumvent direct detection (DD) constraints, as outlined in Refs.\cite{Bhattacharya:2018fus, Bhattacharya:2017sml, Ghorbani:2014qpa, Dutta:2021uxd, Barman:2019oda, Konar:2020wvl, Konar:2020vuu, Sarazin:2021nwo, Ghosh:2021khk, Borah:2021rbx, Mishra:2021ilq, Borah:2022zim}. Singlet leptonic DM has been ruled out due to conflicts between Spin-Independent Direct Detection (SIDD) cross-section and DM relic abundance within a common parameter space\cite{daSilva:2017swg, XENON:2015gkh, XENON:2018voc, Planck:2018vyg, WMAP:2003elm, Bhattacharya:2018fus}. However, singlet fermionic dark matter candidates with suppressed DD cross-sections remain viable. Several studies explore dark sectors involving singlet fermions, extending the SM with additional fermionic and bosonic multiplets. Notably, Ref.\cite{Ghosh:2021khk} extends the SM by introducing a gauged $U(1)_{B-L}$ symmetry. In this framework, the lightest of three exotic singlet fermions, charged under the $U(1)_{B-L}$ gauge group, becomes a dark matter candidate. The model provides a natural suppression of the DD cross-section due to the lack of direct interaction between the dark matter and the Standard Model particles while also offering viable mechanisms for generating neutrino masses and explaining the relic abundance of dark matter.

\par Our work extends SM with a $U(1)_B$ symmetry where exotic fermions are added for gauge anomaly cancellation. The $U(1)_B$ is broken at a scale higher than the electro-weak symmetry breaking~(EWSB) scale, leaving behind a remnant $Z_2$ symmetry under which all the exotic fermions are odd. The $Z_2$ symmetry prevents the decay of the exotic $Z_2$-odd fermions solely into the SM particles, which are even under the $Z_2$. Instead, the exotic fermions can decay into lighter exotic fermions along with one or more SM fermions or scalars. In the absence of any kinematically allowed decay modes for the lightest exotic fermion, it becomes stable and can serve as a cosmologically viable candidate for dark matter. The exotic fermions, being charged under the SM gauge group, have interactions with the SM gauge bosons ($W^\pm/Z/\gamma$). Apart from these interactions, particles in the exotic sector ($Z_2$-odd sector) also interact with the SM particles ($Z_2$-even sector) via interactions mediated by the $Z^\prime$-boson (a massive gauge boson resulting from the broken $U(1)_B$). The production and subsequent decay of these exotic fermions and gauge bosons at collider experiments can give rise to interesting final states. In this work, we have explored dark matter phenomenology in the context of relic density, direct detection of DM, and collider phenomenology within the framework of the Large Hadron Collider experiment.
\par The outline of our work is as follows: In section~\ref{sec:gaugeB}, the model framework, along with the complete Lagrangian, is presented in detail. In section~\ref{sec:scalar}, we discuss the scalar sector of the model along with the bounds on the scalar mixing angle and the complete scalar potential. Section~\ref{sec:exotic} deals with the neutral gauge boson masses and mixings. In section~\ref{sec:mass}, we study the masses and mixing of DM candidates as well as other exotic fermions in the model. Relic density, along with the constraints from direct and indirect detection experiments and the viable parameter space of the model, are discussed in the subsequent sections~\ref{sec:dd},~\ref{sec:relic} and~\ref{sec:idd}. We discuss the collider prospects of the model in section~\ref{sec:collider}. Finally, we concluded in section~\ref{APP:app}.
\section{Model framework}
\label{sec:gaugeB}
The baryonic gauge extension of the SM (here onwards $U(1)_B$ model), where the baryon number is promoted as a local gauge symmetry, is one of the most straightforward extensions beyond the SM. It is based on the following gauge group:
\begin{equation}
 G \equiv SU(3)_C \otimes SU(2)_L \otimes U(1)_Y \otimes U(1)_{B}\,.
\end{equation}
The SM fields and their respective gauge quantum numbers are presented in Table~\ref{table:1}.
\begin{table}[htb!]
\begin{center}
\begin{tabular}{|c|c|c|c|c|}
	\hline
SM Fermions	& $ SU(3)_C$ & $SU(2)_L$ & $U(1)_Y$ & $U(1)_{B}$	\\
	\hline
	\hline
$Q_{L} =  \begin{pmatrix}
             u_L \\ d_L
            \end{pmatrix}$	& $\textbf{3}$ & $\textbf{2}$ & $1/6$ & $1/3$	\\
 $u_R$	& $\textbf{3}$ & $\textbf{1}$ & $2/3$ & $1/3$\\
 $d_R$	& $\textbf{3}$ & $\textbf{1}$ & $-1/3$ & $1/3$\\
$\ell_L =  \begin{pmatrix}
             \nu_L \\ e_L
            \end{pmatrix}$	& $\textbf{1}$ & $\textbf{2}$ & $-1/2$ & $0$	\\
$e_R$	& $\textbf{1}$ & $\textbf{1}$ & $-1$ & $0$	\\
			\hline \hline
\end{tabular}
\end{center}
\vspace{-0.17in}
\caption{Gauge quantum numbers of the SM quarks and leptons under the gauge symmetry $SU(3)_C \otimes SU(2)_L\otimes U(1)_Y \otimes U(1)_{B}$.}
\label{table:1}
\end{table}
A consistent quantum theory must be free from gauge anomalies. So, while introducing a new framework, the cancellation of these anomalies needs to be ensured.
For our $U(1)_{B}$ model, the newly introduced gauge anomalies are the following:
\begin{gather*}
\mathcal{A}[SU(3)^2_C \otimes U(1)_B],  \hspace{.5in} \mathcal{A}[SU(2)^2_L \otimes U(1)_B], \\[0.06in]
\mathcal{A}[U(1)^2_Y \otimes U(1)_B],  \hspace{.5in} \mathcal{A}[U(1)_Y \otimes U(1)^2_B],\\[0.06in]
\mathcal{A}[U(1)^3_B],  \hspace{1in} \mathcal{A}[\text{gravity}^2 \otimes U(1)_B].
\end{gather*}
Four out of these six anomalies vanish and hence do not play any role in our analysis. The non-vanishing gauge anomalies have the following values:
\begin{eqnarray*}
&& \mathcal{A}[SU(2)^2_L \otimes U(1)_B] =\frac{3}{2}, ~~~~~  \mathcal{A}[U(1)^2_Y \otimes U(1)_B] =-\frac{3}{2}.
\end{eqnarray*}
These anomalies can be canceled by adding new fermions charged under the $U(1)_B$ gauge group. Based on earlier works in Ref. \cite{FileviezPerez:2010gw,Chao:2010mp} we extend the SM with a pair of $SU(2)_L$ doublet fermions $(\Psi_{L} \text{ and } \Psi_{R})$, a pair of singlet fermions $\xi_{L}$ and $\xi_{R}$ with non-zero hypercharges, and a pair of singlet fermions $\chi_{L}$ and $\chi_{R}$ with zero-hypercharges. The list of the exotic fermions, along with their gauge quantum numbers, are presented in Table~\ref{table:2}.
\begin{table}
\centering
\begin{tabular}{cccc}
\hline\hline
 & $\text{SU(2)}_L$ & $\text{U(1)}_Y$  & $\text{U(1)}_{B}$\\
 Gauge fields & $\vec{W}_\mu$ & $B_\mu$ & $Z'_{\mu}$\\[1mm]
\hline Fermions \\
(Other than SM) \\
\hline
$\Psi_L = \begin{pmatrix}
           \Psi^+_L \\ \Psi^0_L
          \end{pmatrix}
$ & $\mathbf{2}$ & $\phantom{+}1/2$  & $B_1$\\
$\Psi_R = \begin{pmatrix}
           \Psi^+_R \\ \Psi^0_R
          \end{pmatrix}$ & $\mathbf{2}$ & $1/2$  & $B_2$\\
$\xi_L$ & $\mathbf{1}$ & $1$ & $B_2$\\
$\xi_R$ & $\mathbf{1}$ & $1$ & $B_1$\\
$\chi_L$ & $\mathbf{1}$ & $0$ & $B_2$\\
$\chi_R$ & $\mathbf{1}$ & $0$ & $B_1$\\
\\[1mm] \hline
Scalar fields\\
$H$ & $\mathbf{2}$ & $\phantom{+}1/2$  & $0$ \\
$S$ & $\mathbf{1}$ & $0$ & $B_1-B_2$ \\
\hline\hline
\end{tabular}
\caption{The exotic field content and their transformations under the gauge group $SU(2)_L \otimes U(1)_Y \otimes U(1)_{{B}}$. These fields are chiral in nature and are color singlets, while the electric charge is given by the relation $Q = T^3 + Y$. The detailed framework has been discussed in Refs.~\cite{FileviezPerez:2010gw,Chao:2010mp}.}
\label{table:2}
\end{table}

After introducing the contribution of the new fermions to the gauge anomalies, we are left with the following relations:
\begin{eqnarray}
\nonumber \mathcal{A}[SU(3)^2_c \otimes U(1)_{B}]&=&\frac{1}{2}\cdot \frac{1}{3}\cdot 3\cdot 2+\frac{1}{2}\cdot \frac{-1}{3}\cdot 3\cdot 1+\frac{1}{2}\cdot \frac{-1}{3} \cdot 3\cdot 1+0=0,\\ \nonumber
\mathcal{A}[U(1)_Y \otimes U(1)^2_{B}]&=&2\cdot \frac{1}{2}(B_1^2-B_2^2)+1\cdot(B_2^2-B_1^2)=0,\\ \nonumber
\mathcal{A}[U(1)^3_B] &=&2 \cdot (B_1^3-B_2^3)-(B_1^3-B_2^3)-(B_1^3-B_2^3)=0,\\ \nonumber
\mathcal{A}[\text{gravity}^2 \otimes U(1)_B] &=&2 \cdot (B_1-B_2)-(B_1-B_2)-(B_1-B_2)=0, \\ \nonumber
\mathcal{A}[SU(2)^2_L \otimes U(1)_B] &=&\frac{3}{2}+ \frac{1}{2}(B_1-B_2),\\
\mathcal{A}[U(1)^2_Y \otimes U(1)_B] &=&-\frac{3}{2}-\frac{1}{2}(B_1-B_2).
\label{eq:anom}
\end{eqnarray}

Cancellation of the anomalies requires $B_1-B_2 = -3$, where these $B_1$ and $B_2$ are the $U(1)_{{B}}$ quantum numbers assigned to the new fermions (see Table \ref{table:2}). In our analysis, we have looked into the phenomenology of a Dirac-type DM candidate. This can be achieved for any choice of $B_1$ and $B_2$ other than $B_1 = -B_2$. For convenience, we have fixed their values at $B_1=-1$ and $B_2=2$ (for a detailed discussion on Majorana-type DM candidates in a similar model setting, see Ref. \cite{FileviezPerez:2019jju}). The kinetic and Yukawa part of the Lagrangian for the exotic fermions is given as:

\begin{eqnarray}
 \mathcal{L}&=&\overline{\Psi_{L}}i \slashed{D}\Psi_{L}+\overline{\Psi_{R}}i \slashed{D}\Psi_{R}+\overline{\chi_{L}}i \slashed{D}\chi_{L}+\overline{\chi_{R}}i \slashed{D}\chi_{R} +\overline{\xi_{L}}i \slashed{D}\xi_{L}+\overline{\xi_{R}}i \slashed{D}\xi_{R}\nonumber \\
&&-h_1 \overline{\Psi_{L}} \tilde{H}\xi_R-h_2 \overline{\Psi_{R}} \tilde{H}\xi_L-h_3 \overline{\Psi_{L}} {H}\chi_R -h_4 \overline{\Psi_{R}} {H}\chi_L\nonumber \\ &&-\lambda_{\Psi} \overline{\Psi_{L}} S \Psi_{R} -\lambda_{\xi} \overline{\xi_{L}} \tilde{S} \xi_{R}-\lambda_{\chi} \overline{\chi_{L}} \tilde{S} \chi_{R} .
\label{eq:KYL}
\end{eqnarray}

Here, $H$ is the SM Higgs doublet, and $S$ is the complex scalar singlet (see next section) introduced to break the $U(1)_B$ symmetry. $\Tilde{\phi}~= ~ i \sigma_2 \phi^* (\phi = H,S)$ with $\sigma_2$ the Pauli matrix. 

\subsection{Scalar sector}
\label{sec:scalar}
In our theory, the scalar sector comprises the SM Higgs ($H$) and one complex scalar field $S$ with baryon charge $B_1-B_2=-3$. The second scalar is required to break the $U(1)_B$ symmetry at a higher energy scale. The spontaneous breaking of $U(1)_B$ (also of the EW gauge group) can be triggered through the following scalar potential as discussed in Refs.~\cite{MurguiGalvez:2020mcc,Pruna:2013bma,FileviezPerez:2019jju}:
\bea
-\mathcal{L}^{H,S}_{\rm Mass}&=&
   -\mu^2_{H} H^\dagger H  + \lambda_{H} \big( H^\dagger H \big)^2
-\mu^2_{S} S^\dagger S  +\lambda_{S}\big(S^\dagger S\big)^2 + \lambda_{HS} \big(S^\dagger S\big) \big(H^\dagger H\big).~~~~~
\eea
We can decompose the scalar fields as follows:
\begin{eqnarray}
&&H=\begin{pmatrix}
     h^+ \\
     \frac{1}{\sqrt{2}}(v+\tilde{h})+\frac{i}{\sqrt{2}}A
    \end{pmatrix}
,~~~~S=\frac{1}{\sqrt{2}}(v_B+\tilde{s})+\frac{i}{\sqrt{2}}A_B,
\end{eqnarray}
with $\tilde{h}$ and $\tilde{s}$ being the real part of the neutral components of the scalars $H$ and $S$, respectively. Considering the fact that the VEVs are not zero in the theory's zero-temperature vacuum, which is necessary for the theory to be phenomenologically viable, the minimization conditions are:
\begin{eqnarray}
&&-{\mu_H}^2+\lambda_H v^2+\lambda_{HS}\frac{v_B^2}{2}=0,~~~~~-{\mu_{S}}^2+\lambda_S v_B^2+\lambda_{HS}\frac{v^2}{2}=0.
\label{eq:exofer8}
\end{eqnarray}
For the potential to be bounded from below, we require
\begin{eqnarray}
{ \lambda_H>0, \quad \lambda_S>0 \quad \text{and} \quad\lambda_H\lambda_S-\frac{1}{4}\lambda_{HS}^2>0}.
\label{eq:exofer9}
\end{eqnarray}
Additionally, the perturbativity of the couplings imposes:
\begin{eqnarray}
{ \lambda_H<4 \pi, \quad \lambda_S<4\pi \quad \text{and} \quad \lambda_{HS}<4\pi}.
\label{eq:exofer10}
\end{eqnarray}
After spontaneous symmetry breaking (SSB), the mass matrix for the scalar sector in terms of various scalar couplings and VEVs is given by:
\bea
 \mathcal{M}^2_{HS}
=\left(\begin{matrix}
2\lambda_{H}\,v^2 &  \lambda_{HS}\, v v_B \\
 \lambda_{HS}\, v v_B &2\lambda_{S}\,v_B^2.
 \label{eq:scalMass}
 \end{matrix}\right)
\eea
Upon diagonalizing Eq.~(\ref{eq:scalMass}) the scalar masses can be expressed as:
\bea
&&M^2_{h} = v^2 \lambda_H +v_B^2 \lambda_S  -\sqrt{ (v^2 \lambda_H -v_B^2 \lambda_S)^2+( \lambda_{HS} v v_B)^2,} \nonumber \\
&&M^2_{s} = v^2 \lambda_H +v_B^2 \lambda_S  +\sqrt{ (v^2 \lambda_H -v_B^2 \lambda_S)^2+( \lambda_{HS} v v_B)^2}. \nonumber
\label{eq:exofer11}
\eea
Here, $h$ and $s$ are the physical mass states corresponding to the gauge eigen states $\tilde{h}$ and $\tilde{s}$. The relation between mass and gauge eigenstates can be expressed in a general form:
\begin{eqnarray}
&&h=\cos{\theta} \tilde{h}+\sin{\theta} \tilde{s}, \nonumber \\
&&s=-\sin{\theta} \tilde{h}+\cos{\theta} \tilde{s}.
\label{eq:massScalar}
\end{eqnarray}
Using Eq.~(\ref{eq:scalMass}), we can express the mixing angle~$(\theta)$ as:
\begin{equation}
\tan{(2\theta)}= -\frac{\lambda_{HS} \,v v_B }{\lambda_{S}v_B^2-\lambda_{H}v^2}.
\label{eq:scalartheta}
\end{equation}
Additionally, we can express the mass of physical states $h$ and $s$  in terms of mixing angle~{($\theta$)}~ as:
\bea
\label{ref:phymass}
&&M^2_{h} = 2v^2 \lambda_H  \cos^2\theta +2 v_B^2 \lambda_S \sin^2\theta +\lambda_{HS}\, v v_B   \sin 2\theta, \nonumber \\
&&M^2_{s} = 2v^2 \lambda_H \sin^2\theta + 2v_B^2 \lambda_S \cos^2\theta -\lambda_{HS}\, v v_B  \sin 2\theta.
\label{eq:exofer12}
\eea
Combining equations \ref{eq:scalartheta} and \ref{eq:exofer12}, it is possible to express the scalar couplings in terms of the physical masses and mixing angle. They have the form:
\begin{eqnarray}
&&\lambda_H =\frac{1}{2v^2} \Bigg[{m_h^2 \cos^2{\theta} + m_s^2 \sin^2{\theta}}\Bigg] \leq 4\pi,~~~ \lambda_S = \frac{1}{2v^2_B} \Bigg[{m_h^2 \sin^2{\theta} + m_s^2 \cos^2{\theta}} \Bigg] \leq 4\pi,  \nonumber \\
&&\lambda_{HS}=\frac{1}{2v v_B} \Bigg[{(m_h^2 - m_s^2)}\sin{2\theta}  \Bigg] \leq 4\pi.
\label{eq:exofer13}
\end{eqnarray}

\begin{figure}
\centering
\includegraphics[width=0.78\textwidth]{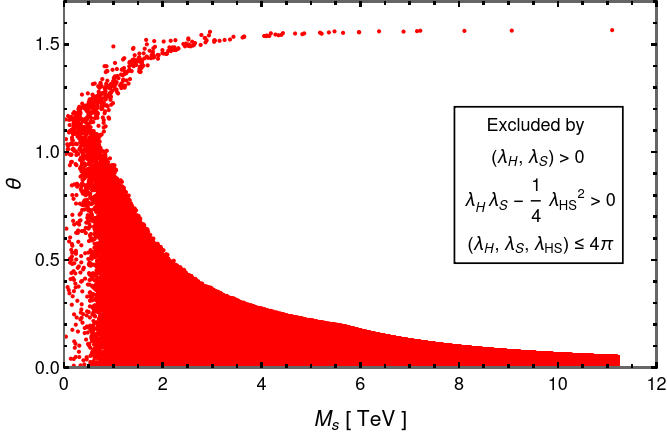}
\caption{The available parameter space in the plane of $M_s -\theta$ is shaded in red, constrained from the conditions in Eq.~(\ref{eq:exofer9}) and (\ref{eq:exofer10}).}
\label{fig:fig1}
\end{figure}

Taking into account the bounded from below and perturbativity constraints, we plot the available parameter space in the $M_s$ Vs $\theta$ plane for a fixed value of $v_B$ = 5 TeV in Figure~\ref{fig:fig1}. From this figure, it is straightforward to determine the upper bound on the new scalar mass as a function of the mixing angle $(\theta)$. Interestingly, we see an overall upper bound of 11.25 TeV on the mass of the exotic scalar $S$ in the $v_B$ = 5 TeV scenario.

\subsection{Gauge sector}
\label{sec:exotic}
The $U(1)_B$ symmetry has an associated gauge boson $Z'$, which acquires non-zero mass after spontaneous breaking of the $U(1)_B$ symmetry. The SSB of $U(1)_B$ symmetry occurs at some higher energy scale by the vacuum expectation value~(VEV) of the exotic scalar field $S$ ($v_B$), which is charged under $U(1)_B$. The exotic scalar field $S$, being a singlet under the SM $SU(2)_L$ and $U(1)_Y$ gauge groups, does not contribute to electroweak symmetry breaking (EWSB). After the EWSB (mediated by the VEV of the SM Higgs doublet H), the SM gauge bosons and fermions acquire their mass. The masses of all the neutral gauge bosons in our theory are given as follows:
\begin{eqnarray}
 && M_A=0, \nonumber \\
 && M_Z =\frac{1}{2} v\sqrt{g^2+g'^2},  \nonumber \\
 &&  M_{Z'}=3 {g_B} {v_B},
 \label{eq:3gbvb}
\end{eqnarray}
where $g_B$ is the gauge coupling associated with $U(1)_B$.
The masses of charged gauge bosons $W^{\pm}$ have their usual SM values~\cite{Halzen:1984mc}. While the SM leptons are not charged under $U(1)_B$, both the SM quarks and the exotic fermions are (see Table \ref{table:1} and Table \ref{table:2}). Consequently, when kinematically allowed, the $U(1)_B$ gauge boson ($Z^\prime$) can decay into either a quark-antiquark pair or a pair of exotic fermions. The corresponding decay widths of $Z^\prime$ are as follows \cite{Duerr:2014wra}:
\begin{align}
  \Gamma ( Z' \rightarrow \bar{q} q ) &= \frac{g_B^2}{36 \pi} M_{Z'} \left(1 - \frac{4 M_q^2}{M_{Z'}^2} \right)^{\frac{1}{2}} \left(1 + \frac{2 M_q^2}{M_{Z'}^2} \right), \\
   \Gamma ( Z' \rightarrow \bar{\chi} \chi ) &= \frac{g_B^2 M_{Z'}}{24 \pi} \left(1 - \frac{4 M_\chi^2}{M_{Z'}^2} \right)^{\frac{1}{2}}  \left[ \left(B_1^2 + B_2^2\right) \left( 1- \frac{M_\chi^2}{ M_{Z'}^2}\right) + 6 B_1 B_2 \frac{M_\chi^2}{ M_{Z'}^2}\right].
\end{align}
Here, $B_1$ and $B_2$ are the leptophobic charges assigned to exotic scalars as given in Table~\ref{table:2} with $B_1-B_2=-3$.

\subsection{Masses and Mixing of Exotic Fermions}
\label{sec:mass}
In Section \ref{sec:gaugeB}, we have discussed the new fermions in our theory required for the cancellation of gauge anomalies along with the Lagrangian describing their interactions. After the spontaneous breaking of $U(1)_B$ and electroweak symmetry, the mass terms for the exotic fermions resulting from Eq.~(\ref{eq:KYL}) are given by,
\begin{eqnarray*}
\mathcal{L}^B_{NF}&=&[-h_1 \overline{\Psi}_{L}^+ \frac{v}{\sqrt{2}}\xi_R^+ -h_2 \overline{\Psi}_{R}^+ \frac{v}{\sqrt{2}}\xi_L^+ -h_3 \overline{\Psi}_{L}^0 \frac{v}{\sqrt{2}}\chi_R^0-h_4 \overline{\Psi}_{R}^0 \frac{v}{\sqrt{2}}\chi_L^0 -\lambda_{\Psi} \overline{\Psi}_{L}^0 \frac{v_B}{\sqrt{2}} \Psi_{R}^0\nonumber \\ && -\lambda_{\Psi} \overline{\Psi}_{L}^+ \frac{v_B}{\sqrt{2}} \Psi_{R}^+ -\lambda_{\xi} \overline{\xi}^+_{L} \frac{v_B}{\sqrt{2}} \xi^+_{R}-\lambda_{\chi} \overline{\chi}_{L}^0 \frac{v_B}{\sqrt{2}} \chi_{R}^0+h.c]
\end{eqnarray*}
This could be re-expressed in terms of mass matrices of the exotic fermions as:
\begin{eqnarray}
\mathcal{L}^B_{NF} &=&
\overline{\left(\begin{matrix} \Psi_L^+ &\xi_{L}^+
\end{matrix}\right)}
{\left(\begin{matrix}
M_{\Psi} & M_2 \\
M_1 & M_{\xi}
\end{matrix}\right)}
{\left(\begin{matrix}
  \Psi_{R}^+ \\ \xi_R^+
\end{matrix}\right)}  +
 \overline{\left(\begin{matrix}
\chi_L^0  & \Psi_{L}^0
\end{matrix}\right)}
{\left(\begin{matrix}
M_{\chi} & M_4 \\
M_3 & M_{\Psi}
\end{matrix}\right)}
{\left(\begin{matrix}
 \chi_R^0 \\  \Psi_{R}^0
\end{matrix}\right)} +h.c.
\label{eq:nflag1}
\end{eqnarray}
The definitions of various matrix elements are listed below,
\begin{equation}
\begin{split}
M_{\Psi}=\frac{\lambda_{\Psi} v_B}{\sqrt{2}},\quad M_{\xi}=\frac{\lambda_{\xi} v_B}{\sqrt{2}}, \quad M_{\chi}=\frac{\lambda_{\chi} v_B}{\sqrt{2}}, \quad M_{1}=\frac{h_1 v}{\sqrt{2}},\\
 \quad M_{2}=\frac{h_2 v}{\sqrt{2}}, \quad M_{3}=\frac{h_3 v}{\sqrt{2}}, \quad M_{4}=\frac{h_4 v}{\sqrt{2}}.~~~~~~~~~~~~
\end{split}
\label{eq:matdef}
\end{equation}
For the neutral exotic fermions, we can express the physical mass basis $\left(\begin{matrix}
 \Psi_{1} &&  \Psi_{2}
\end{matrix}\right)^T_{L,R}$ in terms of the unphysical flavor basis $\left(\begin{matrix}
 \chi^0 && \Psi^0
\end{matrix}\right)^T_{L, R}$ by diagonalizing the second term of Eq.~(\ref{eq:nflag1}) by bi-unitary transformation:
\bea
 \left(\begin{matrix}
 \Psi_{1L} \\ \Psi_{2L}
\end{matrix}\right)
=\mathcal V \left(\begin{matrix}
\chi^0_L \\  \Psi^0_L
\end{matrix}\right)=\left(\begin{matrix}
 \cos\theta_{L}  & \sin\theta_{L} \\
 -\sin\theta_{L}  & \cos\theta_{L}
 \end{matrix}\right)
\left(\begin{matrix}
\chi^0_L \\  \Psi^0_L
\end{matrix}\right),
\label{eq:exofer1}
\eea
\bea
 \left(\begin{matrix}
 \Psi_{1R} \\ \Psi_{2R}
\end{matrix}\right)
=\mathcal V \left(\begin{matrix}
\chi^0_R \\  \Psi^0_R
\end{matrix}\right)=\left(\begin{matrix}
 \cos\theta_{R}  & \sin\theta_{R} \\
 -\sin\theta_{R}  & \cos\theta_{R}
 \end{matrix}\right)
\left(\begin{matrix}
\chi^0_R \\  \Psi^0_R
\end{matrix}\right),
\label{eq:exofer2}
\eea
where $\theta_L$ and $\theta_R$ are the mixing angles in the left and right sectors. From Eqs.~(\ref{eq:exofer1}) and~(\ref{eq:exofer2}), we can define two Dirac fermions $\Psi_1=\Psi_{1L}+\Psi_{1R}$, $\Psi_2=\Psi_{2L}+\Psi_{2R}$, where the lighter one among them can be a viable DM candidate. We have,
\begin{eqnarray}
\Psi_1=\cos{\theta_L}\chi_{L}^0+\sin{\theta_L}\Psi_{L}^0+\cos{\theta_R}\chi_{R}^0+\sin{\theta_R}\Psi_{R}^0, \nonumber \\
\Psi_2=-\sin{\theta_L}\chi_{L}^0+\cos{\theta_L}\Psi_{L}^0-\sin{\theta_R}\chi_{R}^0+\cos{\theta_R}\Psi_{R}^0,
\label{eq:exofer3}
\end{eqnarray}
In our analysis, we adhere to a simplified scenario with $h_3=h_4$ that leads to $\theta_L=\theta_R=\theta_{\text{DM}}$. Under this assumption Eq.~(\ref{eq:exofer3}) takes the form:
\begin{eqnarray}
\Psi_1=\cos{\theta_{\text{DM}}}(\chi_{L}^0+\chi_{R}^0)+\sin{\theta_{\text{DM}}}(\Psi_{L}^0+\Psi_{R}^0), \nonumber \\
\Psi_2=-\sin{\theta_{\text{DM}}}(\chi_{L}^0+\chi_{R}^0)+\cos{\theta_{\text{DM}}}(\Psi_{L}^0+\Psi_{R}^0).
\label{eq:exofer4}
\end{eqnarray}
where the mixing angle $\theta_{\text{DM}}$ is given by the relation,
\begin{eqnarray}\label{ref:mixang}
\tan{2\theta_{\text{DM}}}= - \frac{M_4+M_3}{M_\Psi-M_{\chi}} \equiv -\left(\frac{h_3+h_4}{\lambda_\Psi-\lambda_\chi}\right)\frac{v}{v_B}.
\end{eqnarray}
Assuming,  $m\simeq {M_4+M_3} $, in the small angle limit~($\text{i.e. }\sin\theta_{\text{DM}} \rightarrow \theta_{\text{DM}}$ which is true for $v_B\gg v$), the mass eigenvalues of the physical states $\Psi_1$ and $\Psi_2$ can be expressed as below:
\begin{eqnarray}
M_{\Psi_1} &\simeq& M_{\chi}+\frac{m}{2}\sin{2\theta_{\text{DM}}} \equiv  M_{\chi}-\frac{m^2}{2(M_{\Psi}-M_{\chi})} , \nonumber \\
M_{\Psi_2} &\simeq& M_{\Psi}-\frac{m}{2}_{}\sin{2\theta_{\text{DM}}} \equiv  M_{\Psi}+\frac{m^2}{2(M_\Psi-M_{\chi})} .
\label{eq:DMmass}
\end{eqnarray}
From Eq.~(\ref{eq:DMmass}), it can be seen that in the limit $m \ll m _{\chi} < m_\Psi$, $\Psi_1$ becomes the lightest Dirac fermion and hence the DM candidate.

We may repeat similar calculations for the charged sector fields given in Eq.~(\ref{eq:nflag1}). The final form of mass eigenstates in the charged sector looks like:
\begin{eqnarray}
M_{\Psi_{P_1}} &\simeq& M_{\Psi}-\frac{(M_1+M_2)}{2}\sin{2\theta_{p}}\equiv  M_{\Psi}+\frac{2(M_1+M_2)^2}{(M_\Psi-M_{\xi})},\nonumber \\
M_{\Psi_{P_2}} &\simeq& M_{\xi}+\frac{(M_1+M_2)}{2}\sin{2\theta_{p}}, \equiv  M_{\xi}-\frac{2(M_1+M_2)^2}{(M_\Psi-M_{\xi})},
\label{eq:cflag1}
\end{eqnarray}
with the mixing angle within the charged sector given by,
\begin{equation}
\tan{2\theta_{p}}=-\frac{(M_1+M_2)}{(M_{\Psi}-M_{\xi})} \equiv -\left(\frac{h_1+h_2}{\lambda_\Psi-\lambda_\xi}\right)\frac{v}{v_B}.
\label{eq:cflag2}
\end{equation}
For the purpose of this work, the charged sector mixing angle is taken to be zero, i.e., $\theta_p=0$ in Eq.~(\ref{eq:cflag2}), which is obtained by setting the Yukawa couplings $h_1=h_2=0$. \\

In summary, The neutral component of the doublet fermions $\Psi_{L,R}$ and the neutral singlet fermions $\chi_{L,R}$ combine to form two Dirac fermions ($\Psi_{1,2}$) providing an adequate framework for singlet and doublet fermionic dark matter. The difference between our analysis and previously studied singlet-doublet fermionic dark matter~\cite{Bhattacharya:2018fus} is that all the mass terms in our model are dynamically generated via the spontaneous breaking of $U(1)_B$ gauge symmetry by the non-zero vev of the new scalar field~$(S)$. Additionally, the mixing angle between the doublet and singlet components of the DM helps evade direct detection constraints usually present for a typical WIMP-type dark matter.

The interaction Lagrangian terms for exotic fermions with the Standard Model (SM) gauge bosons and the new gauge boson ($Z'$), denoted as $\mathcal{L}^{EF}_{int}$, are presented in the mass basis of $\Psi_1$, $\Psi_2$, $\Psi_{P_1}$, and $\Psi_{P_2}$ in Appendix~\ref{app:intLag}.
\section{Phenomenology}
\label{sec:cons}
Now that we have introduced the model, we are sufficiently prepared to delve into dark matter and collider phenomenology. This entails examining the constraints imposed on the parameter space of this model by direct and indirect dark matter detection experiments, the measured value of the DM relic density, and collider experiments such as the ongoing Large Hadron Collider (LHC). The phenomenology of the $U(1)_B$ model depends on the following physical parameters:
\bea\label{parms}
\{~M_{\Psi_1}, M_{Z'},~\Delta M(\Psi_2,\Psi_1) , ~\sin\theta_{\text{DM}}, ~v_B \}
\eea
Here, $M_{\Psi_1}$ is the DM mass, $M_{Z'}$ is the mass of leptophobic gauge mediator, $\Delta M(\Psi_2,\Psi_1)$ is the mass difference between the neutral exotic fermions, $\theta_{\text{DM}} $ is the DM mixing angle, and $v_B$ is the VEV responsible for the spontaneous breaking of $U(1)_B$ gauge symmetry.

\subsection{Dark Matter Phenomenology}
Following the spontaneous breaking of $U(1)_B$ and the electroweak symmetry, the $U(1)_B$ model yields two Dirac fermions ($\Psi_1$ and $\Psi_2$), which emerge from the mixing of the neutral components of the exotic doublets and neutral exotic singlets, specifically $\Psi_L^0$, $\Psi_R^{0}$, $\chi_L$, and $\chi_R$ (refer to Table \ref{table:2} and the discussion in section \ref{sec:mass}). The lightest of these two Dirac fermions could potentially serve as a viable candidate for dark matter, subject to meeting the constraints imposed by the dark matter direct and indirect detection experiments, as well as the measured value for dark matter relic density. These constraints will be further elaborated upon in the subsequent sections.


\subsection{Direct Detection Analysis}
\label{sec:dd}
In this section, we discuss the constraints that direct-detection experimental data impose on the parameter space of our model. In these experiments, the DM is assumed to interact with the detector via its scattering off the nucleus inside the detector material, potentially causing the nucleus to recoil. Therefore, on the theoretical front, we are required to estimate the scattering cross-section of DM with the nucleons.

In our framework, the DM-nucleon interaction is induced at the tree level by the mediation of SM Higgs~($h$) and $Z$ boson, and also via exotic scalar~$(s)$ and gauge boson~$(Z')$. The Feynman diagrams corresponding to these interactions can be referred from Figure~\ref{fig:fig13} in Appendix \ref{app:Ca}. As our DM candidate can interact with the SM particles both via scalar and vector mediators, and its spin is not an essential factor in this analysis, the relevant nuclear interactions for our work are the spin-independent ones. Thus, we study the spin-independent direct detection~(SIDD) cross-sections to constrain our model parameter space. Before moving to our results, a few comments are in order. Note that the interactions mediated by $s$ and $Z'$ are suppressed by the heavy mass appearing in the propagator. As for the diagrams mediated by the SM $Z$ boson, their contribution can be handled through a proper choice of the mixing angle $(\theta_{\text{DM}})$, which appears in the amplitude because of the singlet-doublet mixing in the DM candidate (See the discussion around Equation \ref{eq:DMmass}). In our analysis, the numerical values of the spin-independent direct detection cross-sections are computed using micrOmegas \cite{Alguero:2023zol}. We have implemented the BSM model in SARAH~\cite{Staub:2008uz} and used SPheno~\cite{Porod:2011nf} to calculate the spectrum files. This information is then fed into micrOmegas \cite{Alguero:2023zol}, allowing us to calculate quantities such as the dark matter direct detection cross-section and relic density. The next section discusses the constraints on the model parameter space from existing DD experiments.\\

\subsubsection{Direct detection bounds}
The $U(1)_B$ model under study features a two-component singlet-doublet fermionic dark matter candidate, with a mixing angle $\theta_{\text{DM}}$ controlling the contributions from the singlet and doublet components. Since a major contribution to the SIDD cross-section comes from the interaction mediated by the SM $Z$ boson, this mixing angle plays a crucial role in the determination of the SIDD cross-section. \\

\begin{figure}
		\centering

		\begin{subfigure}{0.49\textwidth}
			\includegraphics[width=\linewidth]{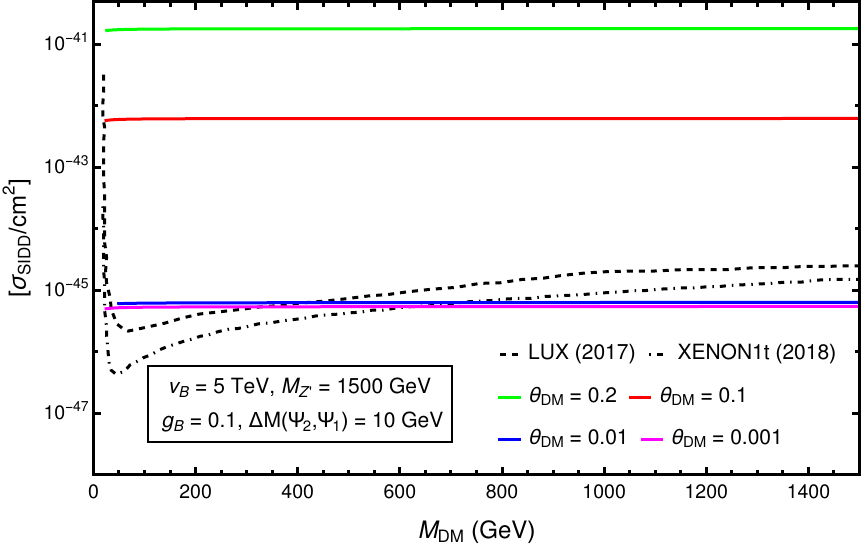}
 			\caption{}
			\label{fig:2sub1}
		\end{subfigure}
		\hfill
		\begin{subfigure}{0.49\textwidth}
			\includegraphics[width=\linewidth]{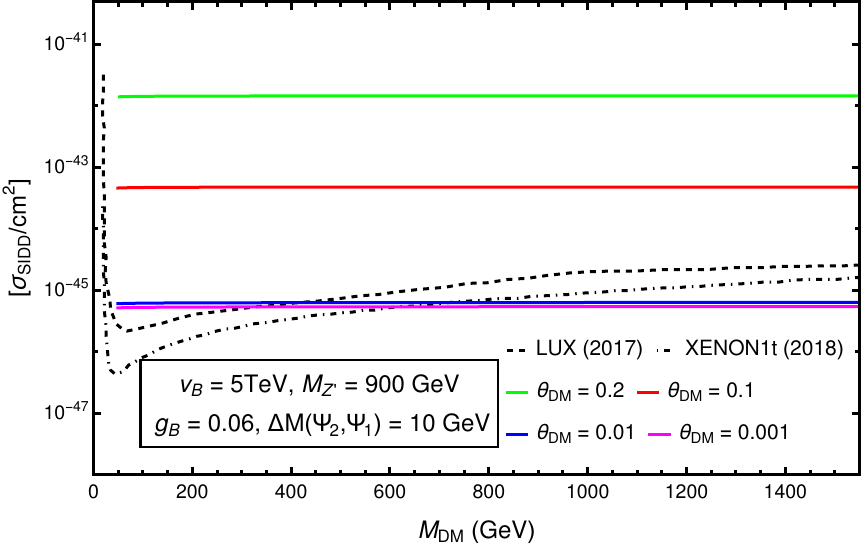}
 			\caption{}
			\label{fig:2sub2}
		\end{subfigure}
		\caption{Spin-independent direct detection~(SIDD) cross-section as a function of DM mass. The limits from LUX~\cite{daSilva:2017swg} and XENON1t~\cite{XENON:2015gkh,XENON:2018voc} experiments have also been plotted~(dashed and dot-dashed black lines, respectively) alongside to compare the results. The graph is plotted for a DM mass range of 0-1500 GeV with the left and right plots for 2 different values of $M_{Z'}$. The various lines corresponds to the different DM mixing angles as mentioned in the figure along with the values of other independent parameters.}
		\label{fig:2DD}
\end{figure}

In Figure~\ref{fig:2DD}, we present the variation of the SIDD cross-section with the DM mass for four different values of $\theta_{\text{DM}}$: 0.001, 0.01, 0.1, and 0.2. For convenience, we have fixed $M_{Z'}$ ($g_B$) at 1500 GeV (0.1) in Figure~\ref{fig:2sub1} and at $900\text{ GeV}$ (0.06) in Figure~\ref{fig:2sub2}. In both figures, the $U(1)_B$ breaking VEV ($v_B$) is fixed at $5\text{ TeV}$, and the mass difference between neutral exotic fermions ($\Delta M(\Psi_2, \Psi_1)$) is set at $10\text{ GeV}$. For comparison, we have also included the upper bounds on the SIDD cross-section from LUX~\cite{daSilva:2017swg} and XENON1T~\cite{XENON:2015gkh,XENON:2018voc} DD experiments. The cross-section strongly depends on $\theta_{\text{DM}}$ up to a minimum threshold. Increasing $\theta_{\text{DM}}$ enhances the doublet component of the DM candidate, which in turn increases the DM-nucleon interaction via $Z$ mediated processes, thus raising the SIDD cross-section. Therefore, there's a constraint on the maximum allowed value $\theta_{\text{DM}}$, beyond which all model parameter space is ruled out by the direct detection data. Figure~\ref{fig:2DD} also demonstrates that, for fixed $v_B$, the behavior of the SIDD cross-section remains unchanged for different $M_{Z'}$ values. This is because the contribution from the $Z^{\prime}$ mediated diagram to the SIDD cross-section is proportional to $g_B^4/M_{Z'}^4 = 1/ v_B^4$ and has no direct dependence on $M_{Z'}$. Additionally, Figure \ref{fig:2DD} shows that DM masses below 600 GeV are excluded by DD constraints for all four values of $\theta_{\text{DM}}$. However, as explored in the subsequent discussion, these bounds can be relaxed with a suitable choice of model parameters.

\begin{figure*}
\centering
\includegraphics[width=0.7\textwidth]{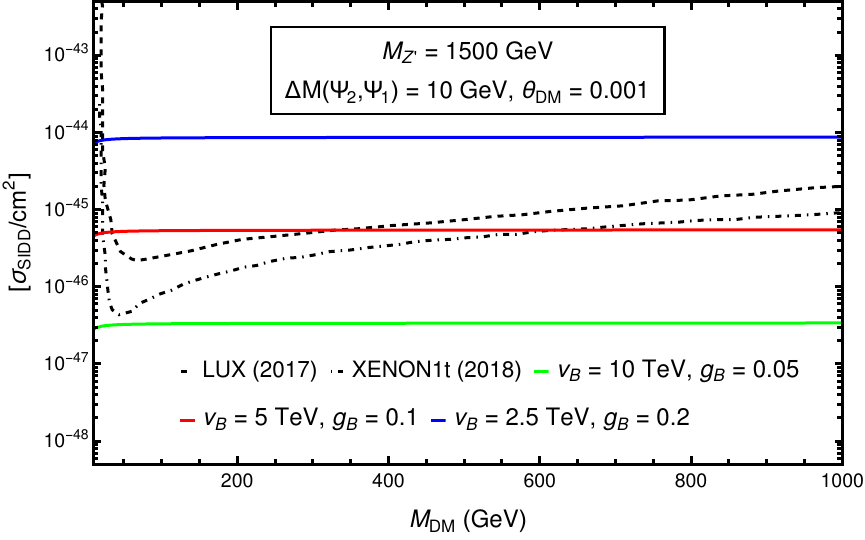}
\caption{SIDD cross-section as a function of  DM mass in GeV for different values of the $U(1)_B$ symmetry VEV, $v_B$. The limits from LUX~\cite{daSilva:2017swg} and XENON1t~\cite{XENON:2015gkh,XENON:2018voc} experiments have also been plotted~(dashed and dot-dashed black lines, respectively) alongside to compare the results. The graph is plotted for a DM mass range of 0-1000 GeV for a fixed value of neutral sector mass splitting, $\Delta M(\Psi_2,\Psi_1)=10\text{ GeV}$, DM mixing angle, $\tan{\theta_{\text{DM}}}=0.001$, and mass of exotic neutral gauge boson, $M_{Z'}=1500\text{ GeV}$.}
\label{fig:fig12n}
\end{figure*}

\begin{figure*}
\centering
\includegraphics[width=0.98\textwidth]{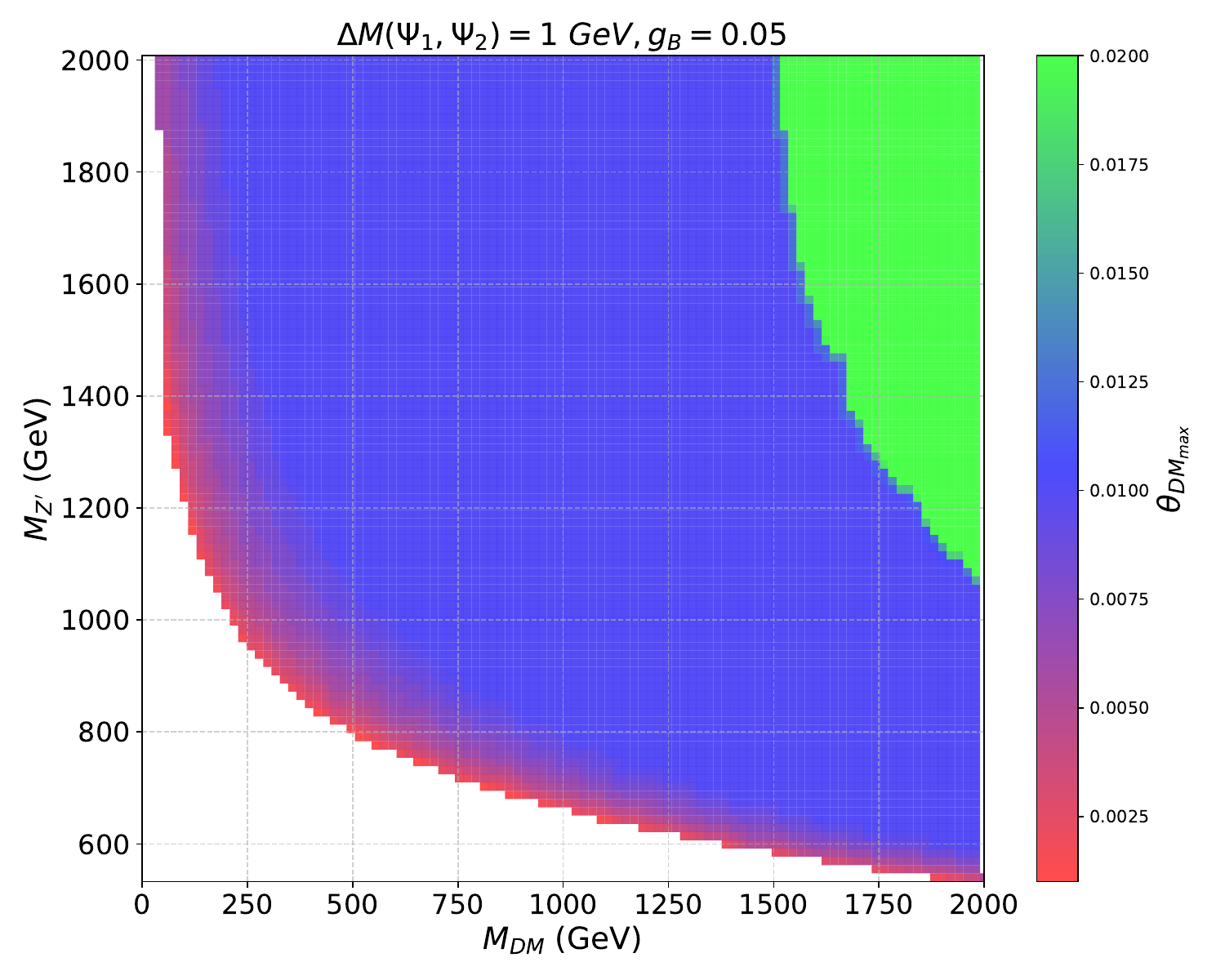}
\caption{A heat map corresponding to the allowed parameter space for our model framework from the direct detection~(DD) experimental data. The color value at each plot point depict the maximum value of DM mixing angle,~$\theta_{\text{DM}}$ that can be used to relax DD constraint for that particular plot point in the $M_{\text{DM}}-M_{Z'}$ plane. We here set $\Delta M(\Psi_2,\Psi_1)=1\text{ GeV}$ and $g_B=0.05$. Values of both the horizontal and vertical axis are taken in GeVs.}
\label{fig:fig14}
\end{figure*}

One key feature to note is the saturation of the direct detection (DD) cross-section beyond a minimum value of $\theta_{\text{DM}}$. This saturation occurs due to a persistent non-zero contribution to the SIDD cross-sections from scalar Higgs ($h$) and exotic gauge boson ($Z'$) mediated processes (see Figure~\ref{fig:fig13}). The $Z'$ mediated contribution can be further suppressed by choosing a high $v_B$, the breaking scale of $U(1)_B$. To illustrate the effects of $v_B$ on the SIDD cross-section, Figure~\ref{fig:fig12n} plots $\sigma_{\text{SIDD}}/\text{cm}^2$ versus $M_{\text{DM}}(\text{GeV})$ for a DM mass range of $10\leq M_{\text{DM}}(\text{GeV})\leq1000$ for three different values of $v_B$: 2.3, 5, and 10 TeV, represented by green blue and red lines, respectively. For this plot, we have fixed $\tan{\theta_{\text{DM}}}$ at 0.001, $M_{Z'}$ at 1500 GeV, the mass splitting between the exotic neutral fermions ($\Delta M(\Psi_2,\Psi_1)$) at 10 GeV, and zero mass splitting between heavier neutral and lighter charged exotic fermions. As expected, the SIDD cross-section decreases with increasing $v_B$. At $v_B = 10\text{ TeV}$, the entire DM mass range is allowed by the DD constraints.\\

To provide a broader perspective on the direct detection allowed parameter space, Figure~\ref{fig:fig14} presents a heatmap for DM mass in the range of $20 \leq M_{\text{DM}} \text{(GeV)} \leq 2000$ and $Z'$ mass in $600 \lesssim M_{Z'} \text{(GeV)} \leq 2000$. The plot shows the allowed values of these parameters from DD constraints provided by LUX~\cite{daSilva:2017swg} and XENON1t~\cite{XENON:2015gkh,XENON:2018voc} experiments. To incorporate the effect of the DM mixing angle ($\theta_{\text{DM}}$), each data point is color-coded to indicate the maximum $\theta_{\text{DM}}$ value that evades DD constraints for that point. For the plot, we have fixed the neutral fermion mass splitting, $\Delta M(\Psi_2, \Psi_1)$, at 1 GeV and the leptophobic gauge coupling strength, $g_B$, at 0.05. Analyzing the plot, we see that for a fixed $M_{\text{DM}}$, the maximum value of $\theta_{\text{DM}}$ allowed by the DD constraints increases with $M_{Z'}$. This is due to the fact that for a fixed $g_B$ with increasing $M_{Z'}$, the contribution from the $Z'$ mediated diagram to the SIDD cross-section decreases. Conversely, as $M_{Z'}$ decreases, the contribution from the $Z'$ mediated diagram increases and can dominate the SIDD cross-section. Under these conditions, the SIDD cross-section becomes independent of $\theta_{\text{DM}}$, and despite taking arbitrarily smaller values of $\theta_{\text{DM}}$ the cross-section can exceed the experimental upper bound. This behavior is depicted in the lower-left corner of Figure \ref{fig:fig14}. As we increase the DM mass for a fixed value of $M_{Z'}$, the upper bound on the SIDD cross-section from the DD experiments gradually increases (see Figure \ref{fig:fig12n}), relaxing the constraints on the model parameter space. Consequently, higher values of $\theta_{\text{DM}}$ become allowed. 

\subsection{Relic Density Calculations}
\label{sec:relic}
In this section, we briefly discuss the relic density of the DM candidate of the $U(1)_B$ model under study. The relic density is influenced by both annihilation and co-annihilation channels, each playing a significant role in determining the final abundance. Relevant Feynman diagrams for these interactions are presented in Figures~\ref{fig:fig2}-\ref{fig:fig6} in Appendix \ref{app:Cb}. In the $U(1)_B$ model under study, the DM relic density is strongly affected by two parameters. The first one is the mass of the $Z'$ boson that determines the DM mass for resonant annihilation. The other one is the mass splitting between the DM and other exotic fermions that significantly impact the co-annihilation channels. We can express the DM relic density as \cite{Griest:1990kh}:
\begin{equation}
	\Omega_{\Psi_1}h'^2=\frac{1.09\times10^9\text{GeV}^{-1}}{g_*^{1/2}M_{PL}}\frac{1}{J(x_f)}
	\label{eq:finalrelic}
\end{equation}
where,
\begin{equation}
	J(x_f)=\int_{x_f}^{\infty}\frac{{\langle\sigma v \rangle}_{eff}}{x^2}dx
	\label{eq:Jf}
\end{equation}
and the effective annihilation cross-section times velocity is denoted by $\langle \sigma v \rangle_{\text{eff}}$ is given as:
\bea
{\langle \sigma v\rangle}_{eff}&&= \frac{g_1^2}{g_{eff}^2} {\langle \sigma v \rangle}_{\overline{\Psi_1}\Psi_1}+\frac{2 g_1 g_2}{g_{eff}^2} {\langle \sigma v \rangle}_{\overline{\Psi_1}\Psi_2}\Big(1+\frac{\Delta M}{M_{\Psi_1}}\Big)^\frac{3}{2}  \nonumber \\
&& \times e^{-x \frac{\Delta M}{M_{\Psi_1}}}+\frac{2 g_1 g_3}{g_{eff}^2} {\langle \sigma v \rangle}_{\overline{\Psi_1}\Psi_{P_i}^+}\Big(1+\frac{\Delta M}{M_{\Psi_1}}\Big)^\frac{3}{2} e^{-x \frac{\Delta M}{M_{\Psi_1}}} \nonumber \\
&& +\frac{2 g_2 g_3}{g_{eff}^2} {\langle \sigma v \rangle}_{\overline{\Psi_2}\Psi_{P_i}^+}\Big(1+\frac{\Delta M}{M_{\Psi_1}}\Big)^3 e^{- 2 x \frac{\Delta M}{M_{\Psi_1}}} \nonumber \\
&& +\frac{g_2^2}{g_{eff}^2} {\langle \sigma v \rangle}_{\overline{\Psi_2}\Psi_2}\Big(1+\frac{\Delta M}{M_{\Psi_1}}\Big)^3 e^{- 2 x \frac{\Delta M}{M_{\Psi_1}}} \nonumber \\
&& +\frac{g_3^2}{g_{eff}^2} {\langle \sigma v \rangle}_{{\Psi_{P_i}^+}\Psi_{P_i}^-}\Big(1+\frac{\Delta M}{M_{\Psi_1}}\Big)^3 e^{- 2 x \frac{\Delta M}{M_{\Psi_1}}},
\label{eq:vf-ann}
\eea

Here, $n_{tot}=n_{\Psi_1}+n_{\Psi_2}+n_{\Psi_P^\pm}$, $\Delta M$ stands for the mass splitting between the exotic neutral fermions~$\Delta M(\Psi_2,\Psi_1)$, $M_{\Psi_1}$ is the DM mass~$(M_{\text{DM}})$, and ${\rm h'}$ is the Hubble parameter. In Eq.~(\ref{eq:vf-ann}), $g_{eff}$ is the effective degree of freedom and is given by:

\begin{equation*}
	g_{eff}=g_1 + g_2 \Big(1+\frac{\Delta M}{M_{\Psi_1}}\Big)^\frac{3}{2} e^{-x \frac{\Delta M}{M_{\Psi_1}}} + g_3\Big(1+\frac{\Delta M}{M_{\Psi_1}}\Big)^\frac{3}{2} e^{-x \frac{\Delta M}{M_{\Psi_1}}} ,
\end{equation*}
Here, $g_1$, $g_2$ and $g_3$ are the degrees of freedom for $\Psi_1$, $\Psi_2$ and $\Psi_{P_i}^{+}$ respectively and $x=x_f={M_{\Psi_1}}/{T_f}$, with $T_f$ being the freeze-out temperature.

In the next section, we discuss the constraints on the $U(1)_B$ model parameter space derived from the measured relic abundance of dark matter as calculated using micrOMEGAs~\cite{Alguero:2023zol}.
\begin{figure}
		\centering

		\begin{subfigure}{0.49\textwidth}
			\includegraphics[width=\linewidth]{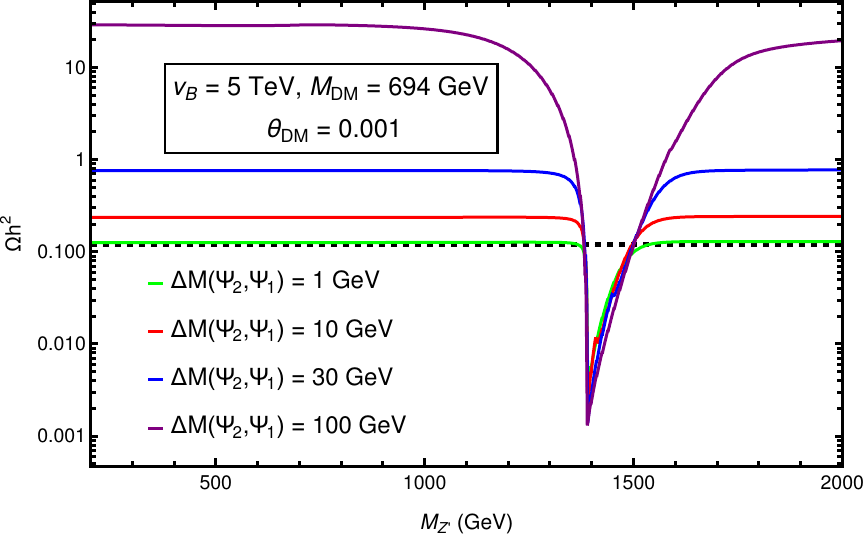}
 			\caption{}
			\label{fig:3sub1}
		\end{subfigure}
		\hfill
		\begin{subfigure}{0.49\textwidth}
			\includegraphics[width=\linewidth]{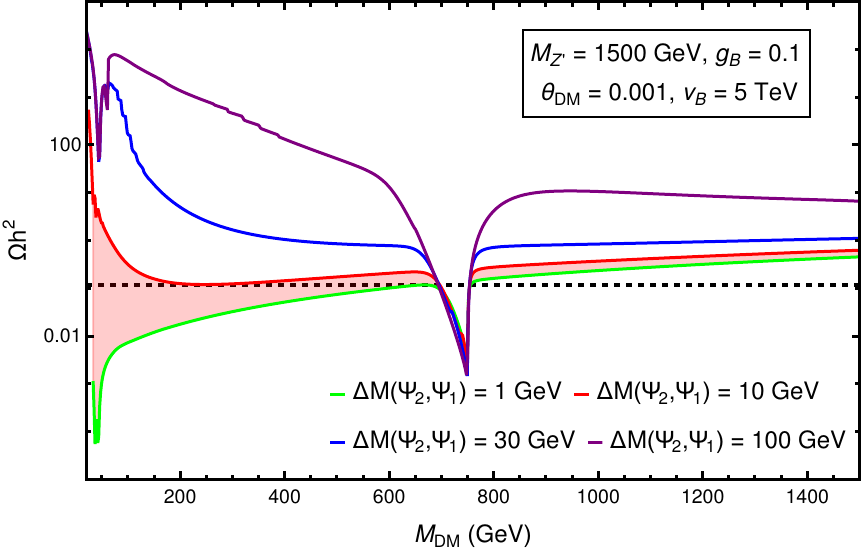}
 			\caption{}
			\label{fig:3sub2}
		\end{subfigure}
		\caption{Relic abundance for the DM as a function of the exotic gauge boson mass~$(M_{Z'})$ in plot~(a) and plot~(b) as a function of DM mass, $M_{\text{DM}}$ . Various colored lines correspond to the various values of mass splitting between the two components of the neutral exotic fermions~$(\Delta M(\Psi_2,\Psi_1))$ in GeV. All the other relevant parameters are mentioned within the plots. The dashed black lines correspond to the observed value of dark matter relic abundance within the $2\sigma$ error~\cite{Planck:2018vyg}.}
		\label{fig:3MZ}
\end{figure}

\begin{figure}
		\centering

		\begin{subfigure}{0.49\textwidth}
			\includegraphics[width=\linewidth]{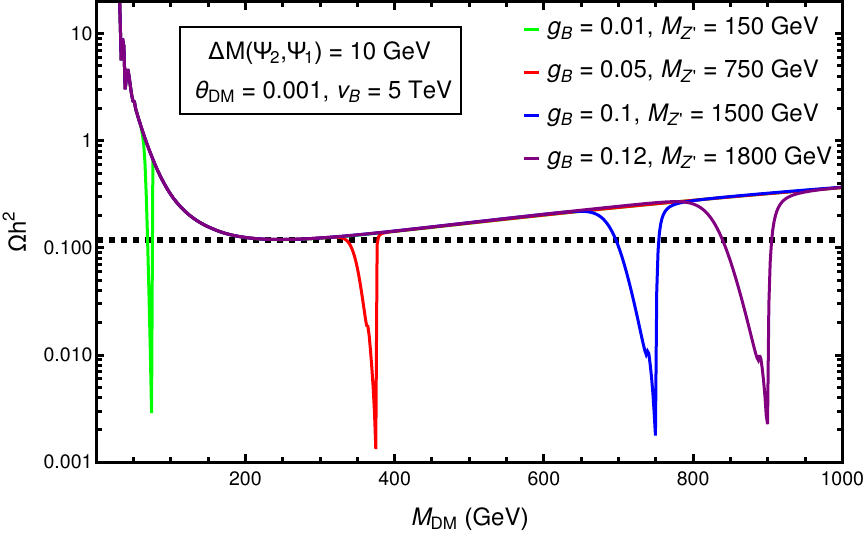}
 			\caption{}
			\label{fig:5sub1}
		\end{subfigure}
		\hfill
		\begin{subfigure}{0.49\textwidth}
			\includegraphics[width=\linewidth]{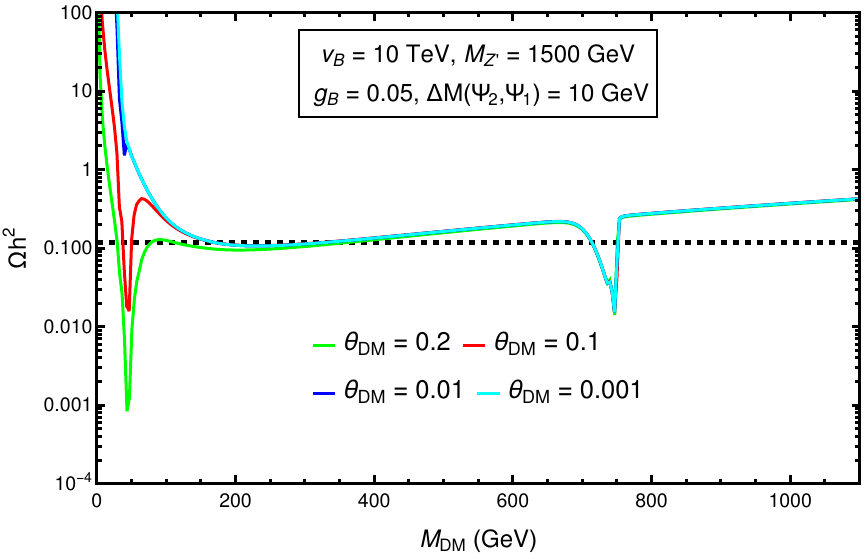}
 			\caption{}
			\label{fig:5sub2}
		\end{subfigure}
		\caption{Two plots for the relic abundance of DM as a function of DM mass, $M_{\text{DM}}$.  Here, both the plots differ by their choices of $U(1)_B$ breaking scale VEV, $v_B$. In plot (a), $v_B$ is set $5$ TeV, whereas in plot (b), $v_B$ is set $10$ TeV. The values of all the other relevant parameters are mentioned within the plots. The dashed black lines correspond to the observed value of dark matter relic abundance within the $2\sigma$ error~\cite{Planck:2018vyg}.}
		\label{fig:5gb}
\end{figure}

\subsubsection{Results and discussions}
\label{subsubsec:RD}
In Figure~\ref{fig:3sub1}, we present the variation of DM relic density with the exotic gauge boson mass~$(M_{Z'})$ for four values of $\Delta M(\psi_1, \psi_2)$: 1, 10, 30, and 100 GeV. For this figure, we have fixed the DM mass at 694 GeV, $v_B$ at 5 TeV, and the mixing angle $\theta_{DM}$ at 0.001. From the plot, we can make the following observations:
\begin{itemize}
	\item For most parts of the parameter space, the relic abundance does not depend on the value of $M_{Z'}$ except at the resonant region where the relic density falls severely due to sudden enhancement in the annihilation cross-section.
	\item The relic density falls with decreasing $\Delta M(\Psi_2,\Psi_1)$. This is due to the increase in the co-annihilation cross-section of the DM as shown in Figures~\ref{fig:fig3}, ~\ref{fig:fig3w}, ~\ref{fig:fig5} \& \ref{fig:fig6} (see Appendix \ref{app:Cb}).
	\item One distinct feature of the plot is the emergence of kinks within the resonance region. They can be accounted for by the contributions of co-annihilation channels (see Figure \ref{fig:fig6}) to the DD cross-section.
\end{itemize}

In Figure~\ref{fig:3sub2}, we present the variation of DM relic density with its mass for different values of $\Delta M(\Psi_2,\Psi_1)$ with a fixed $M_{Z'}$ value of 1500 GeV. In addition to the resonant peak at $M_{Z'}/2$, we see two additional resonant peaks at $M_Z/2$ and $M_h/2$ due to s-channel $Z$ and $h$ mediated interactions, respectively. The relic density decreases with decreasing $\Delta M(\Psi_2,\Psi_1)$ due to an increase in the co-annihilation cross-section. The shaded area in these plots shows the range $\Delta M(\Psi_2,\Psi_1)\in[1-10\text{ GeV}]$ for which we obtain the maximum number of points satisfying the relic density requirement. In figure \ref{fig:5sub1}, we present similar results for four different values of $M_{Z'}$ (150, 750, 1500, and 1800 GeV) at $\Delta M(\Psi_2,\Psi_1)$= 10 GeV. Figure~\ref{fig:5sub2} illustrates the variation of DM relic density with DM mass for four different values of the DM mixing angle($\theta_{\text{DM}} = 0.2, 0.1, 0.01, 0.001$) with a fixed $M_{Z'}$ value of 1500 GeV and corresponds to a mass difference $\Delta M(\Psi_2, \Psi_1)$ of 10 GeV. In this plots, $\Delta M(\Psi_2,\Psi_1)$ being small, the co-annihilation processes dominate, and hence the effect of ${\theta_{\text{DM}}}$ on relic abundance seems suppressed.

\subsection{Indirect Detection Bounds}
\label{sec:idd}
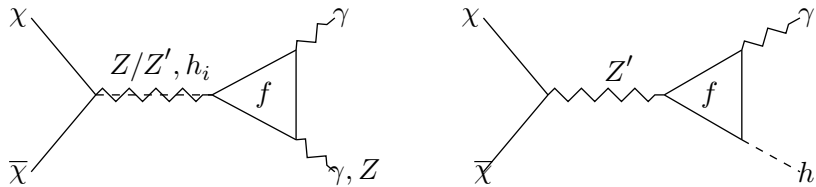
\begin{figure}[H]
\centering
    \begin{tikzpicture}[line width=0.5 pt, scale=0.85]

          \draw[solid] (-1.5,1.2)--(-0.5,0.0);
        \draw[solid] (-1.5,-1.2)--(-0.5,0.0);
         \draw[snake] (-0.5,0.0)--(1.3,0.0);  \draw[dashed] (-0.5,0.0)--(1.3,0.0);
        \draw[solid] (1.3,0.0)--(2.6,0.7);
        \draw[solid] (1.3,0.0)--(2.6,-0.7);
        \draw[solid] (2.6,0.7)--(2.6,-0.7);
        \draw[snake] (2.6,0.7)--(3.2,1.2);
         \draw[snake] (2.6,-0.7)--(3.2,-1.2);
         \node at (-1.7,1.2) {${\chi}$};
         \node at (-1.7,-1.2) {$\overline{\chi}$};
         \node [above] at (0.5,0.05) {$Z/Z', h_i$};

        \node at (2.1,0.0) {$f$};
        \node at (3.3,1.2) {$\gamma$};
        \node at (3.5,-1.2) {$\gamma, Z$};
         \draw[solid] (5.5,1.2)--(6.5,0.0);
        \draw[solid] (5.5,-1.2)--(6.5,0.0);
         \draw[snake] (6.5,0.0)--(8.3,0.0);  

         \draw[solid] (8.3,0.0)--(9.5,0.7);
        \draw[solid] (8.3,0.0)--(9.5,-0.7);
        \draw[solid] (9.5,0.7)--(9.5,-0.7);

         \draw[snake] (9.5,0.7)--(10.4,1.2);
         \draw[dashed] (9.5,-0.7)--(10.4,-1.2);
         \node at (5.3,1.2) {${\chi}$};
         \node at (5.5,-1.2) {$\overline{\chi}$};
         \node [above] at (7.6,0.05) {$Z'$};
          \node at (9.0,0.0) {$f$};
         \node at (10.5,1.2) {$\gamma$};
        \node at (10.5,-1.2) {$h$};
     \end{tikzpicture}
\caption{Annihilation of dark matter particles into gamma rays for the processes ${\chi}\hspace{0.01cm}\overline{\chi} \to \gamma \gamma, \gamma Z, \gamma h$ at loop level for any BSM gauge theory involving an exotic gauge mediator~$(Z')$. Here, $f$ can either be an SM or an exotic fermion present in the theory.}
\label{fig:fig17}
\end{figure}

\begin{figure}[H]
\centering
    \begin{tikzpicture}[line width=0.5 pt, scale=0.85]
          \draw[solid] (-3.0,1.0)--(-1.5,0.0);
        \draw[solid] (-3.0,-1.0)--(-1.5,0.0);
         \draw[snake] (-1.5,0.0)--(0.8,0.0);
        \draw[solid] (0.8,0.0)--(2.5,1.0);
        \draw[snake] (1.59,0.5)--(2.5,0.0);
         \draw[solid] (0.8,0.0)--(2.5,-1.0);
         \node at (-3.3,1.0) {${\chi}$};
         \node at (-3.3,-1.0) {$\overline{\chi}$};
         \node [above] at (0.0,0.05) {$Z'$};
        \node at (2.7,1.0) {$f$};
        \node at (2.7,0.0) {$\gamma$};
        \node at (2.7,-1.0) {$\overline{f}$};
         \draw[solid] (5.0,1.0)--(6.5,0.0);
        \draw[solid] (5.0,-1.0)--(6.5,0.0);
         \draw[snake] (6.5,0.0)--(8.5,0.0);
         \draw[solid] (8.5,0.0)--(10.4,1.0);
         \draw[solid] (8.5,0.0)--(10.4,-1.0);
          \draw[snake] (9.5,-0.5)--(10.4,0.0);

         \node at (4.7,1.0) {${\chi}$};
         \node at (4.7,-1.0) {$\overline{\chi}$};
         \node [above] at (7.4,0.05) {$Z'$};
         \node at (10.5,1.0) {$f$};
        \node at (10.5,-1.0) {$\overline{f}$};
        \node at (10.5,0.0) {$\gamma$};
     \end{tikzpicture}
\caption{Dark matter annihilations that can lead to the production of gamma lines in the final state, also known as final state radiation~(FSR). These processes happen at the tree level. Here, $f$ can either be an SM or an exotic fermion present in the theory.}
\label{fig:fig18}
\end{figure}
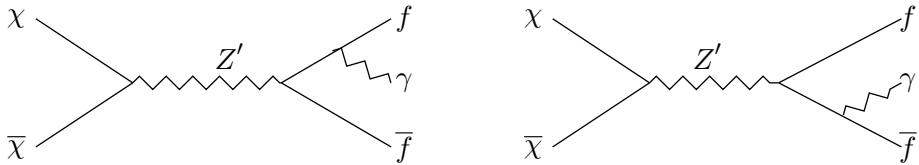

In the experiments dealing with indirect detection, signals from the annihilation of DM into SM particles are sought after. For instance, astrophysical sources of SM particles like positron\cite{Delahaye:2007fr}, antiproton~\cite{Boudaud:2014qra}, gamma ray photons~\cite{Bergstrom:2001jj}, neutrinos~\cite{Arguelles:2017atb} and others~\cite{Duperray:2005si,Donato:2008yx,Dal:2012my,Carlson:2014ssa} are analyzed for any excess in particle fluxes coming from them. One may assume that if an excess in flux of any particle is observed, it can have a DM origin. So, in general, this idea cannot be ruled out. Specifically, the appearance of excess gamma lines from the astrophysical signatures serves as a valuable tool to probe into DM detection, assuming the excess lines come from DM annihilations~\cite{Conrad:2014tla}. However, the production of these distinct energy $\gamma-$lines occurs from a loop-level process ( see Figure. \ref{fig:fig17}) and thus is generally indistinguishable from the tree-level final state radiations (see~Figure \ref{fig:fig18}), except for some specific cases~\cite{Haba:2024vdg,Ciafaloni:2011sa, FileviezPerez:2019rcj}.
\par In our model, the annihilation of DM particles into SM particles can be mediated by the SM $(Z)$ boson, Higgs, or by the exotic bosons like the scalar boson~($s$) and the vector boson $Z'$, as depicted in Figures~\ref{fig:fig2} and \ref{fig:fig4} for the tree-level annihilation processes. Two key criteria must be satisfied for the indirect detection of DM via the production of distinct gamma line signatures. Firstly, the DM annihilation cross-section for the loop-level processes must be high enough to produce a detectable flux of gamma rays. Secondly, the continuous FSR spectrum from other physical processes must be suppressed, ensuring a clean signal with minimal background interference. The DM candidates in our framework are Dirac in nature, so as shown in Ref.~\cite{Duerr:2015wfa}, for no parameter space, we would be able to suppress the contributions of final state radiations~(FSR) in the total photon flux. Thus, the emission of gamma lines, if any, from our DM candidate would overlap and become apparently indistinguishable from the FSR. Thus, an indirect DM analysis can not be performed on our model framework for putting constraints on the parameter space, and neither can it be used to rule out the model.

\subsection{Collider Phenomenology}
\label{sec:collider}
The exotic fermions introduced in Table 2, being charged under the SM gauge group, have gauge interactions with the SM gauge bosons. Moreover, the SM quarks, being charged under the $U(1)_{B}$, interact with the $Z^\prime$ boson. As a result, these TeV-scale exotic fermions and bosons can be produced at colliders, such as the LEP, LHC, and future experiments. The absence of any statistically significant signature of new physics beyond the SM at collider experiments imposes constraints on the parameter space of BSM scenarios. This section is devoted to obtaining bounds on the parameter space of our models, as constrained by the LEP and LHC.

\subsubsection{Decay Widths of the BSM particles}
\label{sec:DW}
The collider signatures of any exotic particles in BSM scenarios crucially depend on the production cross-section, total decay width, and branching ratio of the exotics. While the production cross-section determines the production rate of the signal events, the total decay width and branching ratio determine the topology and characteristics of the final states. Therefore, before discussing the collider bounds on the parameter space of this model, we will first discuss the decay widths of the phenomenologically relevant exotics of the model. In the following, we present the decay width of the  $U(1)_B$ gauge boson $(Z^{\prime})$ into a fermion anti-fermion pair that is charged under the $U(1)_B$:
\begin{equation}
    \Gamma (Z^{\prime} \rightarrow f \Bar{f}) = C_f~ M_{Z^{\prime}} ~[(c_{lf}^2 + c_{rf}^2)(1-r_f)+6~ c_{lf}~c_{rf}~r_f]~\frac{\sqrt{1-4r_f^2}}{24 \pi}
    \label{eqn:dwzp}
\end{equation}
Here,
\begin{align*}
    c_{lq} &= g_B/3\\
    c_{rq} &= g_B/3\\
    c_{l\Psi_1} &= -2~g_B~\cos\theta_{\text{DM}}^2 + g_B~\sin\theta_{\text{DM}}^2\\
    c_{r\Psi_1} &= g_B~\cos\theta_{\text{DM}}^2 - 2~ g_B~\sin\theta_{\text{DM}}^2\\
    c_{l\Psi_2} &= -2 g_B~\sin\theta_{\text{DM}}^2 + g_B~\cos\theta_{\text{DM}}^2\\
    c_{r\Psi_2} &= g_B~\sin\theta_{\text{DM}}^2 - 2~ g_B~\cos\theta_{\text{DM}}^2\\
    c_{l\Psi_{P_1}} &= g_B \\
    c_{r\Psi_{P_1}} &= -2~ g_B
\end{align*}
with $C_f$ being the color factor, q representing all the SM up and down type quarks, and the rest of the symbols carry their usual meaning.

Based on prior experimental bounds~(to be discussed in section~\ref{sec:cp1}), we posit a lower limit of $\ge 500$ GeV for the mass of $Z^{\prime}$. Being leptophobic, $Z^{\prime}$ decays only to the SM quarks and BSM fermions (whenever allowed kinematically). Throughout our analysis, we have assumed the fermion $\Psi_{P_2}$ to have a very high mass (close to 2.5 TeV). When the mass of other BSM fermions (i.e., $\Psi_1, \Psi_2, \rm{and} \Psi_{P_1}$) are small ($\ll M_{Z^\prime}/2$), the $Z^{\prime}$ boson predominantly decays to these BSM fermions, as the decay width into SM quarks is suppressed by the square of their $U(1)_B$ charge. On the other hand, when the masses of the BSM fermions exceed half of $M_{Z^{\prime}}$, the decay into BSM final states is kinematically prohibited, leading $Z^{\prime}$ to decay into SM quark pairs exclusively. In this case, each decay mode has an almost equal branching ratio for $M_Z^{\prime} \ge 500 \text{ GeV}$.

Next, we turn our attention to the particles $\Psi_{P_1}$ and $\Psi_2$. Throughout our analysis, we have assumed no mixing (i.e., $h_1$ = $h_2$ = 0, see the text following Eq.~(\ref{eq:cflag2})) between the charged component of the doublet BSM fermions and the charged singlet BSM fermions. This makes the fermions  $\Psi_{P_1}$ and $\Psi_2$ mass degenerate; hence, one particle's decay into the other is kinematically forbidden. $\Psi_{P_1}$ and $\Psi_2$ can only decay into $\Psi_1$ in association with a $W$ or $Z$ boson. When the mass difference is sufficient, the two-body decay prevails; otherwise, the decay occurs through an off-shell boson into SM fermions. The corresponding three-body decay branching ratios depend on the SM coupling of the quarks and leptons with the $W$ and $Z$ bosons. The explicit forms of two-body decay widths of $\Psi_2$ and $\Psi_{P_1}$ are expressed below as:
\begin{align}
    \Gamma(\Psi_2 \rightarrow \Psi_1 Z) =& \frac{\lambda(M_{\Psi_1}^2,M_{\Psi_2}^2,M_Z^2) ~ |c_{\Psi_1 \Psi_2 Z}|^2}{16\pi M_{\Psi_2}^3} ~ \\&\left(M_{\Psi_1}^2+M_{\Psi_2}^2-2 M_Z^2-6M_{\Psi_1}M_{\Psi_2} + \frac{(M_{\Psi_2}^2-M_{\Psi_1}^2)^2}{M_Z^2} \right) \nonumber
\end{align}
\begin{align}
    \Gamma(\Psi_2 \rightarrow \Psi_1 H) =& \frac{\lambda(M_{\Psi_1}^2,M_{\Psi_2}^2,M_h^2) }{32\pi M_{\Psi_2}^3} ~ \\&\left[ (k_L^2+k_R^2)(M_{\Psi_1}^2+M_{\Psi_2}^2- M_h^2)- 4 k_L k_R M_{\Psi_1} M_{\Psi_2} \right]~~~~~ \nonumber
\end{align}
\begin{align}
    \Gamma(\Psi_{P_1}^{\pm} \rightarrow \Psi_1 W^{\pm}) &= \frac{\lambda(M_{\Psi_1}^2,M_{\Psi_{P_1}}^2,M_W^2) ~ |c_{\Psi_1 \Psi_{P_1} W}|^2}{16\pi M_{\Psi_{P_1}}^3} ~\\ &\left(M_{\Psi_1}^2+M_{\Psi_{P_1}}^2-2 M_W^2-6M_{\Psi_1}M_{\Psi_{P_1}} + \frac{(M_{\Psi_{P_1}}^2-M_{\Psi_1}^2)^2}{M_W^2} \right) \nonumber
\end{align}
Here,
\begin{align}
\lambda(a,b,c) &= \sqrt{a^2+b^2+c^2-2ab-2bc-2ac}\\
c_{\Psi_1 \Psi_2 Z} &= \frac{g}{2 ~\cos\theta_W} ~\sin \theta_{\text{DM}} ~\cos \theta_{\text{DM}}\\
c_{\Psi_1 \Psi_{P_1} W} &= \frac{g}{\sqrt{2}} ~ \sin \theta_{\text{DM}} \\
\begin{split}
k_L &=\frac{1}{\sqrt{2}} [h_3 \sin^2\theta_{\text{DM}} ~\cos\theta - h_4 \cos^2\theta_{\text{DM}} ~\cos\theta \\ &\qquad\qquad\qquad\qquad\qquad + (\lambda_{\chi} - \lambda_{\Psi} ) \sin\theta_{\text{DM}} ~\cos\theta_{\text{DM}} ~\sin\theta ]
\end{split}
\\
\begin{split}
k_R &= \frac{1}{\sqrt{2}} [-h_3 \cos^2\theta_{\text{DM}} ~\cos\theta + h_4 \sin^2\theta_{\text{DM}} ~\cos\theta \\ &\qquad\qquad\qquad\qquad\qquad + (\lambda_{\chi} - \lambda_{\Psi} ) \sin\theta_{\text{DM}} ~\cos\theta_{\text{DM}} ~\sin\theta ]
\end{split}
\end{align}
\noindent The three-body decay widths of $\Psi_2$ and $\Psi_{P_1}$ are presented in Eq.~(\ref{eq:TBD1}) and Eq.~(\ref{eq:TBD2}) in Appendix~\ref{app:TBD}.

While calculating the three-body decay widths, we assumed the final state SM fermions to be massless. Also, note that for the three-body decay of $\Psi_2$, the contributions from the diagrams containing an intermediate Higgs boson are suppressed by the corresponding SM fermion masses and, therefore, are not considered.

\begin{figure}
    \centering
    \includegraphics[width=0.8\columnwidth]{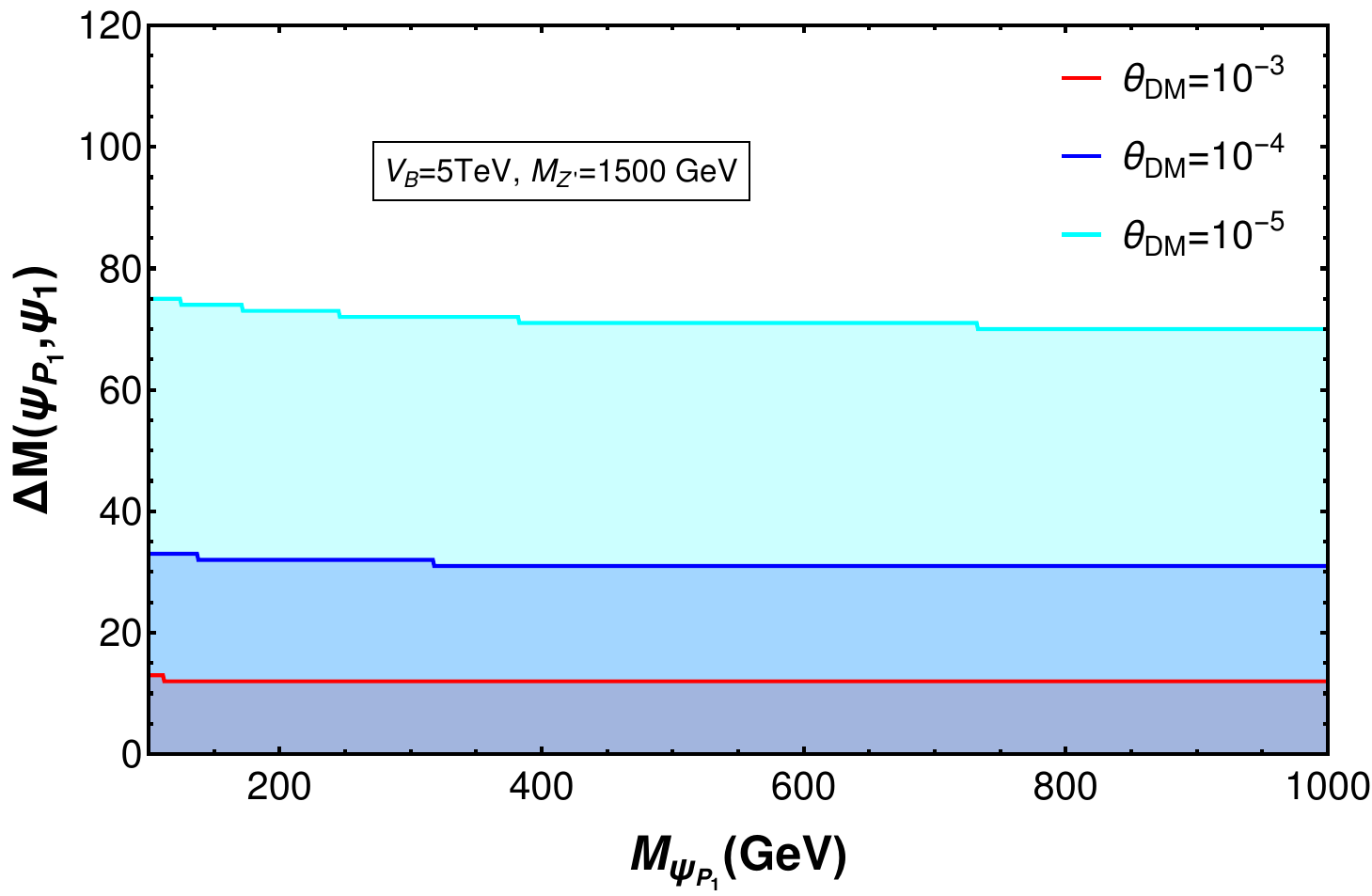}
    \caption{Region of the $M_{\Psi_{P_1}}$ vs $\Delta M(\Psi_{P_1}, \Psi_{1})$ plane where the displaced decay of $\Psi_{P_1}$ prevails. The region in red, blue, and cyan correspond to $\theta_{\text{DM}} = 10^{-3}, 10^{-4}$, and $10^{-5}$ respectively. }
    \label{fig:displaced}
\end{figure}

Our phenomenological analysis relies on the prompt decay of the BSM fermions (See Section \ref{sec:cp2}). However, there is a part of the parameter space associated with very small values of the couplings, where the decay of the BSM fermions is displaced and can also result in decay outside the detector geometry. Since our analysis is not valid for such scenarios, we need to identify that part of the parameter space where the BSM fermions promptly decay. We present the parameter space relevant to our analysis in Figure \ref{fig:displaced}. The shaded region in the figure represents the parameter space where $\Gamma (\Psi_{P_1} \rightarrow Z^{*}(\rightarrow e^- e^+) \Psi_1 ) < 10^{-13}$ and hence favors displaced decay of $\Psi_{P_1}$. The region in red, blue, and cyan correspond to $\theta_{\text{DM}} = 10^{-3}, 10^{-4}$, and $10^{-5}$, respectively. As apparent, a large part of the parameter space allowed by direct detection experiments and relic density constraints favors prompt decay of $\Psi_{P_1}$ and hence is suitable for our collider analysis.

\subsubsection{Constraint from $Z^{\prime}$ search}
\label{sec:cp1}
Here, we discuss the bounds on the $Z^{\prime}$ mass from the ATLAS di-jet analysis~\cite{ATLAS:2018qto}. The analysis targets di-jet events from BSM particles, anticipating distinctive peaks in the invariant mass of the di-jet system, in contrast to the smoothly declining distribution observed in Standard Model (SM) di-jet events. They search for di-jet resonances in the mass range of 450-1800 GeV using 29.3 $fb^{-1}$ of $\sqrt{s}=13 \text{ TeV}$ data collected by the ATLAS detector. The analysis employs a Trigger-Object Level Analysis~(TLA) approach to optimize the selection of trigger-level jets efficiently, minimizing memory usage. Comprehensive details of the analysis, encompassing trigger-level selection, jet calibration, and event criteria, are thoroughly explained in Ref.~\cite{ATLAS:2018qto}. For the sake of completeness, a concise summary is provided here.

Di-jet events are first selected by a first-level~(L1) trigger using a sliding window algorithm. Subsequently, the events undergo further processing via a software-based high-level trigger~(HLT). The HLT retains jets with $p_T > 20 \text{ GeV}$. Two categories of events, one containing at least one trigger-level jet with $E_T > 100 \text{ GeV}$ and the other containing at least one trigger-level jet with $E_T > 75 \text{ GeV}$, are stored for offline analysis. The trigger-level jets then pass through a dedicated calibration procedure to correct the jet's energy and direction. Finally, events with a minimum of two trigger-level jets are selected, with the leading jet requiring $p_T > 220$ or $p_T > 185 \text{ GeV}$ based on the criteria of $E_T > 100 \text{ GeV}$ or $E_T > 75 \text{ GeV}$, respectively. In both cases, the sub-leading jet must have $ p_T > 85 \text{ GeV}$, and all jets must have $|\eta| < 2.8$. Two selection criteria are applied: for the range 700 GeV $< m_{jj} <$ 1800 GeV, events with $|y^{*}| < 0.6$ are chosen, and for $m_{jj} > 450~ GeV$, events from the $E_T > 75$ GeV samples are selected, imposing $|y^{*}| < 0.3$. Here, $y^* = (y_1-y_2)/2$ with $y_1$ and $y_2$ being the rapidities of the  highest and next-to-highest $p_T$ jets. The SM background distribution is determined using a sliding window fit that fits a functional form to the $m_{jj}$ distribution across different bins. The functional form with the least $\chi^2$ across the entire $m_{jj}$ distribution is selected for the final analysis. These results are used to set limits on a leptophobic $Z^{\prime}$ model \cite{Abercrombie:2015wmb} with Dirac fermion dark matter candidate (similar to the BSM model considered in our analysis). The $Z^{\prime}$ is assumed to decay into the SM quarks 100 \% of the time. The model has only two independent parameters $g_q$, the coupling between $Z^{\prime}$ and the SM quarks and $M_{Z^{\prime}}$. The absence of any noteworthy excess in comparison to the SM background leads to a 95 \% credibility-level limit on the di-jet cross-section, which they present as a bound on the ($g_q$, $M_{Z^{\prime}}$) plane.

\begin{figure}
    \centering
    \includegraphics[width=0.8\columnwidth]{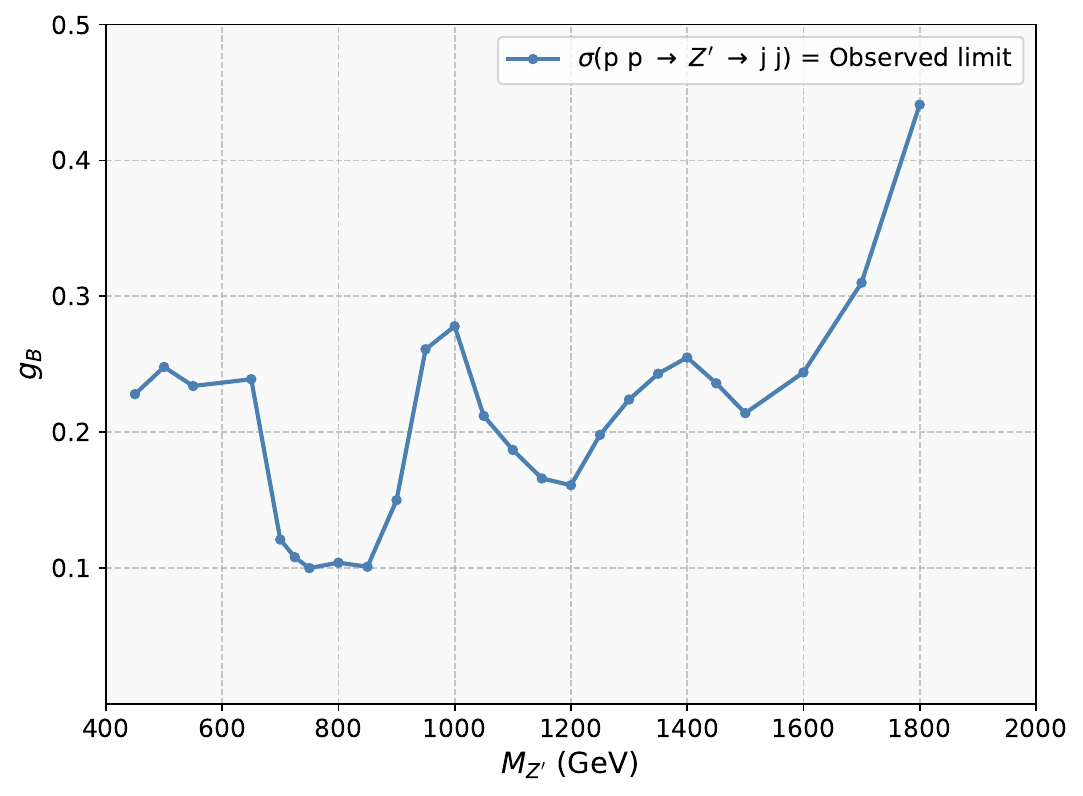}
    \caption{95\% C.L. limits on the $p p \rightarrow Z^{\prime} \rightarrow j j$ cross-section projected onto the $M_{Z^{\prime}}$ vs $g_B$ plane.}
    \label{fig:ca1}
\end{figure}

For our analysis, we followed a conservative approach where we first extracted the bounds on the cross-section for different masses of the $Z^{\prime}$ from the experimental analysis \cite{ATLAS:2018qto}. Later, we used this result to set bound on the leptophobic $U(1)_B$ model parameter space by considering di-jet production through an intermediate $Z^{\prime}$ boson. Note that, for our analysis, we have not considered any detector effects, so the bounds quoted here are conservative at best. We present our results in Figure~\ref{fig:ca1}. For this analysis, we only focus on that part of the parameter space where $M_{\Psi_1}=M_{Z^{\prime}}/2$, $M_{\Psi_2}=M_{\Psi_{P_1}^{\pm}}$, and $\Delta M(\Psi_2,\Psi_1)=10$ GeV. This particular choice of parameters is motivated by the observed relic density of dark matter, a criterion uniquely satisfied within this defined parameter space (as explicated in section~\ref{sec:relic} dedicated to relic density analysis).

\subsubsection{Constraint from Compressed MSSM Search}
\label{sec:cp2}
In this section, we present the bounds on the BSM fermion masses by examining the ATLAS analysis \cite{ATLAS:2019lng}. The above experimental analysis looks into an R-parity conserving compressed MSSM type scenario with 139 $fb^{-1}$ of $\sqrt{s}=13$ TeV pp collision data collected by the ATLAS detector. Four different scenarios are considered in the analysis: one with higgsino-like $\tilde{\chi}_1^0$, $\tilde{\chi}_2^0$, $\tilde{\chi}_1^{\pm}$, one with wino/bino like  $\tilde{\chi}_1^0$, $\tilde{\chi}_2^0$, $\tilde{\chi}_1^{\pm}$, one where they have considered a VBF production mode and one showcasing a slepton NLSP. For our analysis, we only focus on the wino/bino paradigm as it closely resembles the singlet-doublet DM scenario considered in our analysis.

The wino/bino type scenario assumes degenerate $\tilde{\chi}_2^0$ and $ \tilde{\chi}_1^{\pm}$. For this case, \cite{ATLAS:2019lng} considers production of $\tilde{\chi}_2^0 \tilde{\chi}_1^{\pm}$ with the subsequent decay of $\tilde{\chi}_2^0$ into  $\tilde{\chi}_1^0$ and $Z^{*}$ and the decay of $\tilde{\chi}_1^{\pm}$ into  $\tilde{\chi}_1^0$ and $W^{\pm*}$ with a 100\% Branching ratio. Here, the asterisk~$(*)$ over $W^{\pm}$ and Z signifies off-shell states, a consequence of the compressed MSSM spectrum. For the details of the event simulation, object reconstruction, background estimation, etc, we encourage the interested reader to consult Ref.~\cite{ATLAS:2019lng}. The analysis employs the invariant mass of the di-lepton system as the final discriminating variable. The absence of any excess event over the SM background estimation leads to a 95\% C.L. limit on the production cross-section, which they present as a bound on the M($\tilde{\chi}_1^{\pm}$)=M($\tilde{\chi}_2^0$) vs $\Delta M(\tilde{\chi}_2^0,\tilde{\chi}_1^0)$ plane.

Analogous event topology and a mass spectrum are also possible in the case of the leptophobic $U(1)_B$ model considered in our analysis. We can designate $\Psi_1, \Psi_2$ and $\Psi_{P_1}^{\pm}$ as the counterpart of the lightest neutralino, next-to-lightest neutralino, and the lightest chargino, respectively. As discussed in Section~\ref{sec:DW}, when the difference between $M_{\Psi_{P_1}^{\pm}}$ $\approx$  $M_{\Psi_2}$ and $M_{\Psi_1}$ is small, $\Psi_{P_1}^{\pm}$ can only decay into $\Psi_1$ and an off-shell $W^{\pm}$, while $\Psi_2$ can only decay into $\Psi_1$ and an off-shell $Z/h$ boson. However, the contribution of the off-shell $h$-mode to the three body decay width is highly suppressed due to the relatively smaller coupling of $h$ with the SM fermions. It is crucial to note that since $\Psi_1$ and $\Psi_2$ are Dirac fermions, consideration must also be given to their respective anti-particles. (In the subsequent text, a mention of $\Psi_1$ or $\Psi_2$ automatically implies that their corresponding anti-particles are also to be considered for the analysis).

\begin{figure}[H]
    \centering
    \includegraphics[width=0.8\columnwidth]{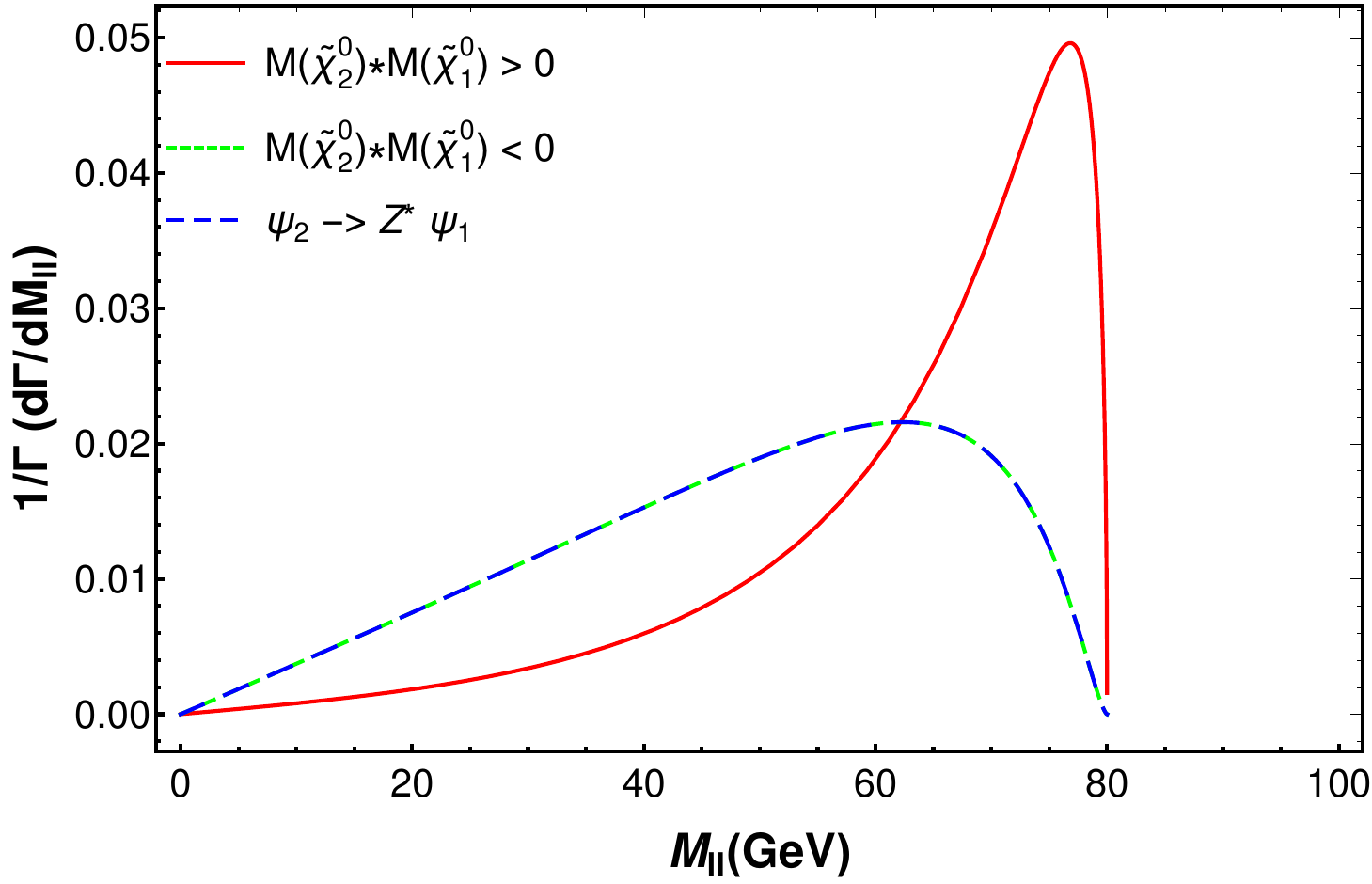}
    \caption{Dilepton invariant mass for the wino/bino SUSY case as well as for the $U(1)_B$ model under study.}
    \label{fig:invmass}
\end{figure}

The relative sign between the mass of the lightest and next-to-lightest neutralino plays a very crucial role in determining the decay kinematics. Therefore, the final experimental bounds also depend on the relative sign of $M(\tilde{\chi}_1^0)$ and $M(\tilde{\chi}_2^0)$. In Ref.~\cite{ATLAS:2019lng}, the cases of $M(\tilde{\chi}_1^0)\times M(\tilde{\chi}_2^0) < 0$ and $M(\tilde{\chi}_1^0)\times M(\tilde{\chi}_2^0) >0$ are considered separately. To determine which of these scenarios better fits our study, we calculate analytically the Dilepton invariant mass from the decay of our BSM fermion $\Psi_2$ and compare it with the corresponding results from Ref.~\cite{ATLAS:2019lng}. We present our results in Figure~\ref{fig:invmass}. For generating the plots we have considered the case of $M(\tilde{\chi}_2^0)=M_{\Psi_2} = 180$ GeV and $M(\tilde{\chi}_1^0)=M_{\Psi_1} = 100$ GeV. The plot shows that the scenario under study resembles the $M(\tilde{\chi}_1^0)\times M(\tilde{\chi}_2^0) < 0$ case, and hence the corresponding experimental bounds are more appropriate for our analysis.

\begin{figure}[H]
    \centering
    \includegraphics[width=0.8\columnwidth]{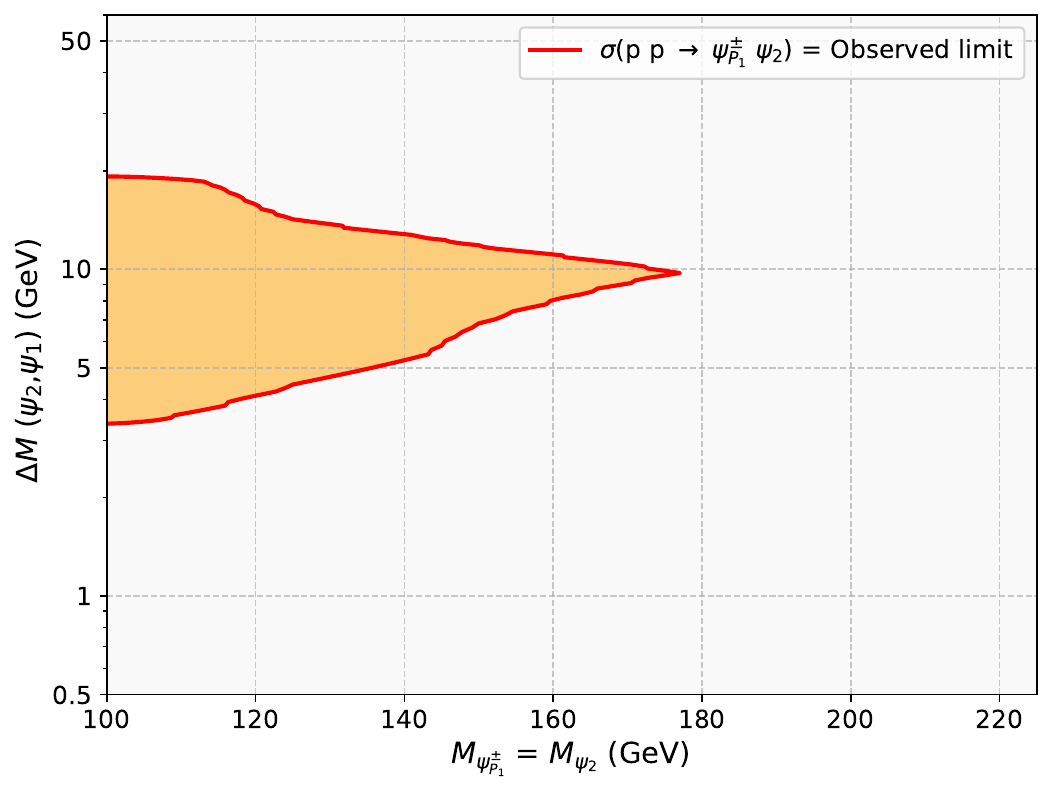}
    \caption{ 95\% C.L. limit on the $\Psi_{P_1}^{\pm} \Psi_2$ production cross-section projected into the $M_{\Psi_{P_1}^{\pm}} = M_{\Psi_2}$ vs $\Delta M(\Psi_2,\Psi_1)$ plane.}
    \label{fig:ca2}
\end{figure}

All these being said, in the limit $M_{\Psi_{P_1}^{\pm}} = M_{\Psi_2}$, the collider production of $\Psi_{P_1}^{\pm}$ $\Psi_2$ shares similar kinematics as the chargino-neutralino case considered above and the corresponding experimental results can be reinterpreted to set bounds on the $U(1)_B$ model parameter space. However, for our analysis, instead of pursuing a dedicated collider analysis, we took a conservative approach. We extracted the 95 \% C.L. bounds on the chargino-neutralino production cross-section from the experimental analysis of Ref.~\cite{ATLAS:2019lng} and used this result to set bound on the production cross-section of $\Psi_{P_1}^{\pm}$ $\Psi_2$ without incorporating any details of detector effect. Subsequently, we project the bounds into the $M_{\Psi_{P_1}^{\pm}} = M_{\Psi_2}$ vs $\Delta M(\Psi_2,\Psi_1)$ plane, see Figure \ref{fig:ca2}. The plot reveals that a substantial portion of the parameter space is still available for further studies.

\section{Conclusion}
\label{sec:conc}
In this work, we have explored a class of minimal leptophobic~$(U(1)_{B})$ gauge extension of the SM, wherein the lightest neutral fermion~(emerging as a singlet-doublet admixture of exotic neutral fermions) added for anomaly cancellation serves as a viable WIMP dark matter candidate. A $Z_2$ discrete symmetry remnant from the breaking of $U(1)_B$ by a TeV-scale scalar $S$ ensures the stability of our DM particles. Recently, there has been a surge in DM models that attempt to offer some breathing space in the direct detection parameter space, which otherwise seems to rule out a purely singlet or a purely doublet type WIMP DM. The approach has been to relax the spin-independent direct detection~(SIDD) cross-section by extending the SM with fermions that can form mixed-state DM candidates offering a weaker interaction with nucleons. Based on this motivation, we here explore the possibility of a singlet-doublet mixed state DM~(tuned by the DM mixing angle parameter, $\theta_{\text{DM}}$) in the leptophobic SM extension and provide a significant parameter space for our DM that can be probed from current and future direct-detection~\cite{XENON:2018voc,LZ:2022lsv} and collider experiments~\cite{ATLAS:2019lng}.

The uniqueness of our work lies in the fact that the exotic fermions added for anomaly cancellation are sufficient to provide a stable DM candidate saved from direct detection experiments, thus keeping the framework minimal, and also the leptophobic gauge Boson $Z'$ paves the way for extra DM annihilation and interaction channels thus offering new physics signatures at colliders to probe. We observe that the SIDD cross-section constraints are circumvented for a singlet-dominated dark matter state, provided the dark matter mass exceeds 600 GeV and the dark matter mixing angle is less than or equal to 0.01. This analysis is conducted with other relevant input parameters held at specific values: $M_{Z'} = 1500$ GeV, $g_B = 0.1$, and $U(1)_B$ VEV $(v_B)$ set at 5 TeV. We further find that with an increased value of $v_B$ i.e. at 10 TeV, the parameter space expands to encompass the entire relevant range of dark matter masses from 1 GeV to 1000 GeV for a leptophobic gauge coupling value of $g_B=0.2$ with $M_{Z'}=1500\text{ GeV}$ and $\theta_{\text{DM}}=0.001$. Within this large parameter space, we show that relic requirements are satisfied easily by tuning the mass splitting in the neutral sector~$\Delta M(\Psi_2,\Psi_1)$ from 1 to 10 GeVs. For completeness, we also investigate the effects of other phenomenological relevant model parameters, such as dark matter mixing angle $\theta_{\text{DM}}$, the mass of the exotic gauge Boson $M_{Z'}$, and the $U(1)_B$ gauge coupling $g_B$, on the relic abundance of the DM candidate.

At the collider front, we have looked into constraints on the model parameter space based on two ATLAS analyses \cite{ATLAS:2018qto,ATLAS:2019lng}. The analysis \cite{ATLAS:2018qto}, studies a generic leptophobic $Z^{\prime}$ vector boson and sets bounds on the $M_{Z^{\prime}}$ Vs $g_q$ (the coupling between the $Z^{\prime}$ and SM quarks) parameter space. We have performed a reinterpretation of the analysis for the $U(1)_B$ model under study for a scenario where $M_{\Psi_1}$ = ($M_{Z^{\prime}}$)/2, $M_{\Psi_2} = {M_{\Psi_{P_1}^{\pm}}}$, and $\Delta$M($\Psi_2,\Psi_1) = 10$ GeV. Our results presented in figure \ref{fig:ca1} show that most of the parameter space allowed by direct detection and relic density data does not have any significant constraints and is available for further exploration. The second experimental analysis \cite{ATLAS:2019lng} looked into an R-parity conserving compressed MSSM type scenario with Bino/Wino like $\Tilde{\chi}_1^0$, $\Tilde{\chi}_2^0$, and $\Tilde{\chi}_1^{\pm}$. The final state topology considered can be achieved in the leptophobic $Z^{\prime}$ model under study and this allows us to reinterpret the experimental analysis to set constraints on the BSM model parameter space. Our results presented in figure \ref{fig:ca2} demonstrate that the search can only exclude a limited part of the parameter space leaving the rest for future experimental analyses to explore.

This work presents promising prospects for future extensions and analysis. For the exotic fermions, we chose a specific combination of leptophobic charges to obtain a Dirac-type DM candidate. It's noteworthy that these charge combinations can be adjusted to generate Majorana or pseudo-Dirac-type dark matter particles, thereby opening up new parameter spaces for exploration. For a Majorana type WIMP DM, there can arise suppressed tree level cross-sections for final state radiations, allowing distinct gamma lines peaks to show up. This can provide potential signatures detectable through indirect searches such as HESS~\cite{HESS:2013rld}, CTA~\cite{CTAConsortium:2017dvg}. Additionally, we are conducting a gravitational wave (GW) analysis for this framework in a separate study, where we demonstrate that the breaking of the $U(1)_B$ symmetry at high energies could impact early universe cosmic bubble formations. The presence of leptophobic scalar $S$ can give rise to noticeable gravitational waves that may be resolved at experiments like  NANOGrav~\cite{NANOGrav:2023hvm}, KARGA~\cite{KAGRA:2019htd} and DECIGO~\cite{Kawamura:2023fwf}. The near-future testability prospects of this framework, both through direct detection setups and complementary GW searches, underscore the phenomenological relevance and significance of our work. More stringent direct searches of a WIMP dark matter by collaborations like LZ~\cite{LZ:2022lsv} in the coming years would serve as an invaluable tool to validate our framework's viability.

\section*{Acknowledgments}
Taramati acknowledges the financial support from the Department of Science and Technology INSPIRE~(DST/INSPIRE Fellowship/IF200289), New Delhi. Utkarsh Patel (UP) would like to express gratitude for the financial support received from the Ministry of Education, Government of India. UP also extends appreciation for the hospitality provided by Dr. Kirtiman Ghosh during his visit to IOP, where a significant part of the analysis has been conducted.\\ [0.2in]
\noindent \begin{Large}\textbf{Appendix}\end{Large}
\appendix
\label{APP:app}
\section{Derivation of DM interaction Lagrangian}
\label{app:intLag}
The most general kinetic term for the DM constituent candidates~$(\Psi\text{ and }\chi)$ can be written as:-
\bea
\mathcal{L}&=&  \{\underbrace{i\overline{\Psi_L} \slashed{D}\Psi_L+i\overline{\Psi_R} \slashed{D}\Psi_R }_{(\mathcal{L}_1)}\}+ \{\underbrace{i\overline{\chi_L} \slashed{D}\chi_L +i\overline{\chi_R} \slashed{D}\chi_R}_{(\mathcal{L}_2)}\}
\label{eq:Tot}
\eea
Here, the exotic fermions~$(\Psi_L, \Psi_R, \chi_L, \chi_R)$ have their interactions with not only the SM gauge Bosons but also with the exotic gauge Boson~$(Z')$. So, firstly we present below the interaction Lagrangian terms for exotic fermions with the SM gauge bosons,~($\mathcal{L}^{EF}_{int}$) in their mass basis representation~($\Psi_1$, $ \Psi_2$, $\Psi_{P_1}$ and $ \Psi_{P_2}$) as:
\bea
\mathcal{L}^{EF}_{int} &=& e_0 \Big[\sin^2\theta_{\text{DM}} \overline{\Psi_1}\gamma^{\mu}A_{\mu}\Psi_1+
\cos^2\theta_{\text{DM}} \overline{\Psi_2}\gamma^{\mu}A_{\mu}\Psi_2
\nonumber \\
&&+
\sin\theta_{\text{DM}} \cos\theta_{\text{DM}} (\overline{\Psi_1}\gamma^{\mu}A_{\mu}\Psi_2+\overline{\Psi_2}\gamma^{\mu}A_{\mu}\Psi_1)\Big]  \nonumber \\
&&+
\Big(\frac{e_0\cos2\theta_W}{2\sin\theta_W \cos\theta_W}\Big)\Big[\sin^2\theta_{\text{DM}} \overline{\Psi_1}\gamma^{\mu}Z_{\mu}\Psi_1+
\cos^2\theta_{\text{DM}} \overline{\Psi_2}\gamma^{\mu}Z_{\mu}\Psi_2
\nonumber \\
&&+
\sin\theta_{\text{DM}} \cos\theta_{\text{DM}} (\overline{\Psi_1}\gamma^{\mu}Z_{\mu}\Psi_2+\overline{\Psi_2}\gamma^{\mu}Z_{\mu}\Psi_1)\Big]
\nonumber \\
&&+\frac{e_0}{\sqrt2\sin\theta_W}\sin\theta_{\text{DM}} \cos\theta_{\text{P}} \overline{\Psi_{P_1}}\gamma^\mu W_\mu^+ \Psi_1 -\frac{e_0}{ \sqrt2\sin\theta_W} \sin\theta_{\text{DM}} \sin\theta_{\text{p}} \overline{\Psi_{P_2}}\gamma^\mu W_\mu^+ \Psi_1  \nonumber \\
&&+\frac{e_0}{\sqrt2\sin\theta_W}\cos\theta_{\text{DM}} \cos\theta_{\text{P}} \overline{\Psi_{P_1}}\gamma^\mu W_\mu^+ \Psi_2 -\frac{e_0}{\sqrt2 \sin\theta_W} \cos\theta_{\text{DM}} \sin\theta_{\text{p}} \overline{\Psi_{P_2}}\gamma^\mu W_\mu^+ \Psi_2  \nonumber \\
&&+\frac{e_0}{\sqrt2\sin\theta_W}\sin\theta_{\text{DM}} \cos\theta_{\text{P}} \overline{\Psi_{1}}\gamma^\mu W_\mu^- \Psi_{P_1} +\frac{e_0}{ \sqrt2\sin\theta_W} \cos\theta_{\text{DM}} \cos\theta_{\text{p}} \overline{\Psi_{2}}\gamma^\mu W_\mu^- \Psi_{P_1}  \nonumber \\
&&-\frac{e_0}{\sqrt2\sin\theta_W}\sin\theta_{\text{DM}} \sin\theta_{\text{P}} \overline{\Psi_{1}}\gamma^\mu W_\mu^- \Psi_{P_2} -\frac{e_0}{ \sqrt2\sin\theta_W} \cos\theta_{\text{DM}} \sin\theta_{\text{p}} \overline{\Psi_{2}}\gamma^\mu W_\mu^- \Psi_{P_2}  \nonumber \\
&&+ \Big({e_0 \cos^2\theta_{\text{p}}}\Big) {\Psi_{P_1}}\gamma^{\mu}A_{\mu}\Psi_{P_1}+ \Big({e_0\sin^2\theta_{\text{p}}}\Big) {\Psi_{P_2}}\gamma^{\mu}A_{\mu}\Psi_{P_2} \nonumber \\
&&- \Big(\frac{e_0\sin2\theta_{\text{p}}}{2}\Big) {\Psi_{P_2}}\gamma^{\mu}A_{\mu}\Psi_{P_1}- \Big(\frac{e_0\sin2\theta_{\text{p}}}{2}\Big) {\Psi_{P_1}}\gamma^{\mu}A_{\mu}\Psi_{P_2} \nonumber \\
&&+ \Big({\Psi_{P_1}}\gamma^{\mu}Z_{\mu}\Psi_{P_1}{\cos^2\theta_{\text{p}}}+{\Psi_{P_2}}\gamma^{\mu}Z_{\mu}\Psi_{P_2}{\sin^2\theta_{\text{p}}}\Big)\Big(e_0 {\cot2\theta_W} \Big)  \nonumber \\
&&- \Big({\Psi_{P_2}}\gamma^{\mu}Z_{\mu}\Psi_{P_1}-{\Psi_{P_1}}\gamma^{\mu}Z_{\mu}\Psi_{P_2}\Big)\Big(e_0 \frac{\cot2\theta_W}{2}{\sin2\theta_{\text{p}}} \Big)
.\label{eq:massint1}
\eea
Now, we also express the interaction terms for the exotic fermions~$(\Psi_L, \Psi_R, \chi_L, \chi_R)$ with the $Z'$ gauge boson,~($\mathcal{L'}^{EF}_{int}$) in their mass basis representation~($\Psi_1$, $ \Psi_2$, $\Psi_{P_1}$ and $ \Psi_{P_2}$). For reader's clarity, we have divided the total Lagrangian term in Eq.~($\ref{eq:Tot}$) into $\mathcal{L}_1$ and $\mathcal{L}_2$ for $\Psi$ and $\chi$ interactions, respectively. So, considering first the $\Psi$ part, we have
\begin{eqnarray*}
\mathcal{L}_1 &=& \{i\overline{\Psi_L} \slashed{D}\Psi_L+i\overline{\Psi_R} \slashed{D}\Psi_R \} \nonumber \\
 &&= -ig_B  \overline{\begin{pmatrix}
\Psi_L^+ & \Psi_L^0
\end{pmatrix}}
B_1 \slashed{Z'}
{\begin{pmatrix}
\Psi_L^+ \\ \Psi_L^0
\end{pmatrix}}-ig_B  \overline{\begin{pmatrix}
\Psi_R^+ & \Psi_R^0
\end{pmatrix}}
B_2 \slashed{Z'}
{\begin{pmatrix}
\Psi_R^+ \\ \Psi_R^0
\end{pmatrix}} \nonumber \\
&&=-i g_B\overline{\Psi_L^+} \slashed{Z'} B_1 \Psi_L^+ -i g_B\overline{\Psi_R^+} \slashed{Z'} B_2 \Psi_R^+\{\underbrace{-i g_B\overline{\Psi_L^0} \slashed{Z'} B_1 \Psi_L^0 -i g_B\overline{\Psi_R^0} \slashed{Z'} B_2 \Psi_R^0}_{(\mathcal{L}_1^0)}\}
\end{eqnarray*}
Here, we take only the extra term due to the extra gauge boson~$Z'$
Using $\Psi_L=\Psi_{1L}+\Psi_{2L}$,~$\Psi_R=\Psi_{1R}+\Psi_{2R}$ and by assuming $\theta_L=\theta_R=\theta_{\text{DM}}$, we have $(\mathcal{L}_1^0)$ re-expressed as,
\begin{eqnarray*}
\mathcal{L}_1^0&=& -i g_B(\sin{\theta_L}\overline{\Psi_{1L}}+\cos{\theta_L}\overline{\Psi_{2L}})\slashed{Z'} B_1 (\sin{\theta_L}{\Psi_{1L}}+\cos{\theta_L}{\Psi_{2L}}) \nonumber\\&&~~~~-i g_B(\sin{\theta_R}\overline{\Psi_{1R}}+\cos{\theta_R}\overline{\Psi_{2R}})\slashed{Z'} B_2 (\sin{\theta_R}{\Psi_{1R}}+\cos{\theta_R}{\Psi_{2R}})\nonumber \\[.1in]
 &=&-ig_B B_1[\sin^2{\theta}_L({\overline{\Psi_{1L}} \slashed{Z'}  \Psi_{1L}})+\sin{\theta}_L\cos{\theta}_L({\overline{\Psi_{2L}} \slashed{Z'}  \Psi_{1L}}) \nonumber \\ &&~~~~+\sin{\theta}_L\cos{\theta}_L({\overline{\Psi_{1L}} \slashed{Z'}  \Psi_{2L}})+\cos^2{\theta}_L({\overline{\Psi_{2L}} \slashed{Z'}  \Psi_{2L}})] \nonumber \\
  &&~~~~-ig_B B_2[\sin^2{\theta}_R({\overline{\Psi_{1R}} \slashed{Z'}  \Psi_{1R}})+\sin{\theta}_R\cos{\theta}_R({\overline{\Psi_{2R}} \slashed{Z'}  \Psi_{1R}}) \nonumber \\ &&~~~~+\sin{\theta}_R\cos{\theta}_R({\overline{\Psi_{1R}} \slashed{Z'}  \Psi_{2R}})+\cos^2{\theta}_R({\overline{\Psi_{2R}} \slashed{Z'}  \Psi_{2R}})]
\nonumber \\[.1in]
&=&-ig_B B_1[\sin^2{\theta}_L({\overline{\Psi_{1}} \slashed{Z'} P_L  \Psi_{1}})+\sin{\theta}_L\cos{\theta}_L({\overline{\Psi_{2}} \slashed{Z'} P_L \Psi_{1}}) \nonumber \\ &&~~~~+\sin{\theta}_L\cos{\theta}_L({\overline{\Psi_{1}} \slashed{Z'} P_L \Psi_{2}})+\cos^2{\theta}_L({\overline{\Psi_{2}} \slashed{Z'} P_L \Psi_{2}})] \nonumber \\
  &&~~~~-ig_B B_2[\sin^2{\theta}_R({\overline{\Psi_{1}} \slashed{Z'} P_R \Psi_{1}})+\sin{\theta}_R\cos{\theta}_R({\overline{\Psi_{2}} \slashed{Z'} P_R  \Psi_{1}}) \nonumber \\ &&~~~~+\sin{\theta}_R\cos{\theta}_R({\overline{\Psi_{1}} \slashed{Z'} P_R   \Psi_{2}})+\cos^2{\theta}_R({\overline{\Psi_{2}} \slashed{Z'} P_R  \Psi_{2}})] \nonumber \\[.1in]
  &=& -ig_B \sin^2{\theta_{\text{DM}}}{\overline{\Psi_{1}} \slashed{Z'}  }(B_1 P_L+B_2 P_R)\Psi_{1}-ig_B \cos^2{\theta_{\text{DM}}}{\overline{\Psi_{2}} \slashed{Z'}  }(B_1 P_L+B_2 P_R)\Psi_{2} \nonumber \\
  &&~~~~-ig_B \sin{\theta_{\text{DM}}}\cos{\theta_{\text{DM}}}{\overline{\Psi_{2}} \slashed{Z'}  }(B_1 P_L+B_2 P_R)\Psi_{1} \nonumber \\
  &&~~~~ -ig_B \sin{\theta_{\text{DM}}}\cos{\theta_{\text{DM}}}{\overline{\Psi_{1}} \slashed{Z'}  }(B_1 P_L+B_2 P_R)\Psi_{2}
\end{eqnarray*}
Similarly, considering only the $\chi$ part, we have
\begin{eqnarray*}
\mathcal{L}_2 &=& \{i\overline{\chi_L} \slashed{D}\chi_L+i\overline{\chi_R} \slashed{D}\chi_R\} \nonumber \\[.1in]
&=& -ig_B \overline{\chi_L}^0
B_2 \slashed{Z'} \chi_L^0
-ig_B \overline{\chi_R}^0
B_1 \slashed{Z'} \chi_R^0 \nonumber \\[.1in]
&=&-i g_B\overline{\chi_L^0} \slashed{Z'} B_2 \chi_L^0 -i g_B\overline{\chi_R^0} \slashed{Z'} B_1 \chi_R^0 \nonumber\\[.1in]
&=& -i g_B(\cos{\theta_L}\overline{\Psi_{1L}}-\sin{\theta_L}\overline{\Psi_{2L}})\slashed{Z'} B_2 (\cos{\theta_L}{\Psi_{1L}}-\sin{\theta_L}{\Psi_{2L}}) \nonumber\\&&~~~~-i g_B(\cos{\theta_R}\overline{\Psi_{1R}}-\sin{\theta_R}\overline{\Psi_{2R}})\slashed{Z'} B_1 (\cos{\theta_R}{\Psi_{1R}}-\sin{\theta_R}{\Psi_{2R}})\nonumber \\[.1in]
&=&-ig_B B_2[\cos^2{\theta}_L({\overline{\Psi_{1L}} \slashed{Z'}  \Psi_{1L}})-\sin{\theta}_L\cos{\theta}_L({\overline{\Psi_{2L}} \slashed{Z'}  \Psi_{1L}}) \nonumber \\ &&~~~~-\sin{\theta}_L\cos{\theta}_L({\overline{\Psi_{1L}} \slashed{Z'}  \Psi_{2L}})+\sin^2{\theta}_L({\overline{\Psi_{2L}} \slashed{Z'}  \Psi_{2L}})] \nonumber \\
 &&~~~~-ig_B B_1[\cos^2{\theta}_R({\overline{\Psi_{1R}} \slashed{Z'}  \Psi_{1R}})-\sin{\theta}_R\cos{\theta}_R({\overline{\Psi_{2R}} \slashed{Z'}  \Psi_{1R}}) \nonumber \\ &&~~~~-\sin{\theta}_R\cos{\theta}_R({\overline{\Psi_{1R}} \slashed{Z'}  \Psi_{2R}})+\sin^2{\theta}_R({\overline{\Psi_{2R}} \slashed{Z'}  \Psi_{2R}})]
\nonumber \\[.1in]
&=&-ig_B B_1[\cos^2{\theta}_L({\overline{\Psi_{1}} \slashed{Z'} P_L  \Psi_{1}})-\sin{\theta}_L\cos{\theta}_L({\overline{\Psi_{2}} \slashed{Z'} P_L \Psi_{1}}) \nonumber \\ &&~~~~-\sin{\theta}_L\cos{\theta}_L({\overline{\Psi_{1}} \slashed{Z'} P_L \Psi_{2}})+\sin^2{\theta}_L({\overline{\Psi_{2}} \slashed{Z'} P_L \Psi_{2}})] \nonumber \\
  &&~~~~-ig_B B_2[\cos^2{\theta}_R({\overline{\Psi_{1}} \slashed{Z'} P_R \Psi_{1}})-\sin{\theta}_R\cos{\theta}_R({\overline{\Psi_{2}} \slashed{Z'} P_R  \Psi_{1}}) \nonumber \\ &&~~~~-\sin{\theta}_R\cos{\theta}_R({\overline{\Psi_{1}} \slashed{Z'} P_R   \Psi_{2}})+\sin^2{\theta}_R({\overline{\Psi_{2}} \slashed{Z'} P_R  \Psi_{2}})]  \nonumber \\[.1in]
  &=& -ig_B \cos^2{\theta_{\text{DM}}}{\overline{\Psi_{1}} \slashed{Z'}  }(B_2 P_L+B_1 P_R)\Psi_{1}-ig_B \sin^2{\theta_{\text{DM}}}{\overline{\Psi_{2}} \slashed{Z'}  }(B_2 P_L+B_1 P_R)\Psi_{2} \nonumber \\
  &&~~~~+ig_B \sin{\theta_{\text{DM}}}\cos{\theta_{\text{DM}}}{\overline{\Psi_{2}} \slashed{Z'}  }(B_2 P_L+B_1 P_R)\Psi_{1} \nonumber \\
  &&~~~~+ig_B \sin{\theta_{\text{DM}}}\cos{\theta_{\text{DM}}}{\overline{\Psi_{1}} \slashed{Z'}  }(B_2 P_L+B_1 P_R)\Psi_{2}
\end{eqnarray*}
Putting in the final expression for $\mathcal{L}_1^0$ in $\mathcal{L}_1$ and then combining the results of $\mathcal{L}_1$ and $\mathcal{L}_2$, the total interaction Lagrangian for the vector portal of DM states is given as:
\begin{eqnarray*}
 \mathcal{L}&=&-ig_B \cos^2{\theta_{\text{DM}}}({\overline{\Psi_{1}} \slashed{Z'}  })(B_2 P_L+B_1 P_R)\Psi_{1}-ig_B \sin^2{\theta_{\text{DM}}}({\overline{\Psi_{2}} \slashed{Z'}  })(B_2 P_L+B_1 P_R)\Psi_{2} \nonumber \\
  &&~~~~+ig_B \sin{\theta_{\text{DM}}}\cos{\theta_{\text{DM}}}({\overline{\Psi_{2}} \slashed{Z'}  })(B_2 P_L+B_1 P_R)\Psi_{1} \nonumber \\
  &&~~~~+ig_B \sin{\theta_{\text{DM}}}\cos{\theta_{\text{DM}}}({\overline{\Psi_{1}} \slashed{Z'}  })(B_2 P_L+B_1 P_R)\Psi_{2} \nonumber \\
  &&~~~~-ig_B \sin^2{\theta_{\text{DM}}}({\overline{\Psi_{1}} \slashed{Z'}  })(B_1 P_L+B_2 P_R)\Psi_{1}-ig_B \cos^2{\theta_{\text{DM}}}({\overline{\Psi_{2}} \slashed{Z'}  })(B_1 P_L+B_2 P_R)\Psi_{2} \nonumber \\
  &&~~~~-ig_B \sin{\theta_{\text{DM}}}\cos{\theta_{\text{DM}}}({\overline{\Psi_{2}} \slashed{Z'}  })(B_1 P_L+B_2 P_R)\Psi_{1} \nonumber \\
  &&~~~~-ig_B \sin{\theta_{\text{DM}}}\cos{\theta_{\text{DM}}}({\overline{\Psi_{1}} \slashed{Z'}  })(B_1 P_L+B_2 P_R)\Psi_{2}\nonumber \\&&~~~~-i g_B\overline{\Psi_L^+} \slashed{Z'} B_1 \Psi_L^+ -i g_B\overline{\Psi_R^+} \slashed{Z'} B_2 \Psi_R^+ \nonumber \\[.1in]
  &=&-ig_B ({\overline{\Psi_{1}} \slashed{Z'}  })[(B_2 P_L+B_1 P_R)\cos^2{\theta_{\text{DM}}}+(B_1 P_L+B_2 P_R)\sin^2{\theta_{\text{DM}}}]\Psi_{1}\nonumber \\
  &&~~~~-ig_B ({\overline{\Psi_{2}} \slashed{Z'}  })[(B_2 P_L+B_1 P_R)\sin^2{\theta_{\text{DM}}}+(B_1 P_L+B_2 P_R)\cos^2{\theta_{\text{DM}}}]\Psi_{2}\nonumber \\
  &&~~~~-ig_B ({\overline{\Psi_{2}} \slashed{Z'}  })\sin{\theta_{\text{DM}}}\cos{\theta_{\text{DM}}}[-(B_2 P_L+B_1 P_R)+(B_1 P_L+B_2 P_R)]\Psi_{1}
  \nonumber \\
  &&~~~~-ig_B ({\overline{\Psi_{1}} \slashed{Z'}  })\sin{\theta_{\text{DM}}}\cos{\theta_{\text{DM}}}[-(B_2 P_L+B_1 P_R)+(B_1 P_L+B_2 P_R)]\Psi_{2}\nonumber \\&&~~~~-i g_B\overline{\Psi_L^+} \slashed{Z'} B_1 \Psi_L^+ -i g_B\overline{\Psi_R^+} \slashed{Z'} B_2 \Psi_R^+
\end{eqnarray*}
Now, inserting the free parameter values $B_1=-1$ and $B_2=2$, we have
\begin{eqnarray*}
\mathcal{L}&=&-ig_B ({\overline{\Psi_{1}} \slashed{Z'}  })[ P_L(2\cos^2{\theta_{\text{DM}}}-\sin^2{\theta_{\text{DM}}})+P_R(2\sin^2{\theta_{\text{DM}}}-\cos^2{\theta_{\text{DM}}})]\Psi_{1}\nonumber \\
&&~~-ig_B ({\overline{\Psi_{2}} \slashed{Z'}  })[ P_L(2\sin^2{\theta_{\text{DM}}}-\cos^2{\theta_{\text{DM}}})+P_R(2\cos^2{\theta_{\text{DM}}}-\sin^2{\theta_{\text{DM}}})]\Psi_{2}\nonumber \\
&&~~-ig_B ({\overline{\Psi_{2}} \slashed{Z'}  })\sin{\theta_{\text{DM}}}\cos{\theta_{\text{DM}}}[3P_R- 3P_L]\Psi_{1}\nonumber \\
&&~~-ig_B ({\overline{\Psi_{1}} \slashed{Z'}  })\sin{\theta_{\text{DM}}}\cos{\theta_{\text{DM}}}[3P_R- 3P_L]\Psi_{2} \nonumber \\&&~~-i g_B\overline{\Psi_L^+} \slashed{Z'} B_1 \Psi_L^+ -i g_B\overline{\Psi_R^+} \slashed{Z'} B_2 \Psi_R^+ \nonumber \\[.1in]
&=&-ig_B ({\overline{\Psi_{1}} \slashed{Z'}  })[ P_L(2-3\sin^2{\theta_{\text{DM}}})+P_R(3\sin^2{\theta_{\text{DM}}}-1)]\Psi_{1}\nonumber \\
&&~~-ig_B ({\overline{\Psi_{2}} \slashed{Z'}  })[ P_L(3\sin^2{\theta_{\text{DM}}}-1)+P_R(2-3\sin^2{\theta_{\text{DM}}})]\Psi_{2}\nonumber \\
&&~~-ig_B ({\overline{\Psi_{2}} \slashed{Z'} })\sin{2\theta_{\text{DM}}}\frac{3(P_R- P_L)}{2} \Psi_{1}
 -ig_B ({\overline{\Psi_{1}} \slashed{Z'}  })\sin{2\theta_{\text{DM}}}\frac{3(P_R- P_L)}{2}\Psi_{2} \nonumber \\&&~~-i g_B\overline{\Psi_L^+} \slashed{Z'} B_1 \Psi_L^+ -i g_B\overline{\Psi_R^+} \slashed{Z'} B_2 \Psi_R^+
\end{eqnarray*}
So, the interaction terms between the exotic DM neutral states and $Z'$ in the mass eigenbasis are expressed as below:
\begin{eqnarray*}
 \mathcal{L}&=&-ig_B ({\overline{\Psi_{1}} \slashed{Z'}  })[ P_L(2-3\sin^2{\theta_{\text{DM}}})+P_R(3\sin^2{\theta_{\text{DM}}}-1)]\Psi_{1}\nonumber \\
&&~~-ig_B ({\overline{\Psi_{2}} \slashed{Z'}  })[ P_L(3\sin^2{\theta_{\text{DM}}}-1)+P_R(2-3\sin^2{\theta_{\text{DM}}})]\Psi_{2}\nonumber \\
&&~~-ig_B ({\overline{\Psi_{2}} \slashed{Z'}  })\sin{2\theta_{\text{DM}}}\frac{3(P_R- P_L)}{2}\Psi_{1}
 -ig_B ({\overline{\Psi_{1}} \slashed{Z'}  })\sin{2\theta_{\text{DM}}}\frac{3(P_R- P_L)}{2}\Psi_{2} \nonumber \\&&~~-i g_B\overline{\Psi_L^+} \slashed{Z'} B_1 \Psi_L^+ -i g_B\overline{\Psi_R^+} \slashed{Z'} B_2 \Psi_R^+
\end{eqnarray*}

\section{Three body decays}
\label{app:B}
The three-body decay width of $\Psi_2$ state into $\Psi_1 f \Bar{f}$ is expressed as follows:
\label{app:TBD}
\begin{align}
    \frac{d\Gamma (\Psi_2 \rightarrow \Psi_1 f \Bar{f})}{dQ} =& \frac{8 C_f c_{\Psi}}{(\mu + M)^3} Q \frac{\sqrt{Q^4-Q^2(\mu^2+M^2)+(\mu M)^2}}{(Q^2-M_Z^2)^2}\\
    & (-4Q^4+Q^2(M^2+\mu^2) - 3Q^2(M^2-\mu^2)+2(\mu M)^2) \nonumber
     \label{eq:TBD1}
\end{align}
Here, $C_f$ is the associated color factor and,
\begin{align}
    c_{\Psi} &= \frac{\pi^2 ~g^2~ c_{\Psi_1 \Psi_2 Z}^2 (g_v^2+g_A^2)}{6~ \cos {\theta_W^2} ~(2 \pi)^5}\\
    \mu &= M_{\Psi_2} - M_{\Psi_1}\\
     M &= M_{\Psi_2} + M_{\Psi_1}\\
     g_v &= \frac{1}{2} T^3_f - Q_f \sin^2{\theta_W}  \\
     g_A &= \frac{1}{2} T^3_f
\end{align}

\noindent Similarly, the three-body decay widths of $\Psi_{P_1}$ into $\Psi_{1}f\Bar{f^{\prime}}$ is expressed as below,
\begin{align}
    \frac{d\Gamma (\Psi_{P_1} \rightarrow \Psi_1 f \Bar{f^{\prime}})}{dQ} =& \frac{8 C_f c_{\chi}}{(\mu + M)^3} Q \frac{\sqrt{Q^4-Q^2(\mu^2+M^2)+(\mu M)^2}}{(Q^2-M_W^2)^2}\\
    & (-4Q^4+Q^2(M^2+\mu^2) - 3Q^2(M^2-\mu^2)+2(\mu M)^2) \nonumber
    \label{eq:TBD2}
\end{align}
 Here, $C_f$ is the associated color factor and,
\begin{align}
    c_{\chi} &= \frac{\pi^2 g^2 c_{\Psi_1 \Psi_{P_1} W}^2}{24~(2 \pi)^5}\\
    \mu &= M_{\Psi_{P_1}} - M_{\Psi_1}\\
     M &= M_{\Psi_{P_1}} + M_{\Psi_1}
\end{align}
\section{Feynman Diagrams}
\label{app:C}
\subsection{Feynman diagrams relevant with direct detection}
\label{app:Ca}
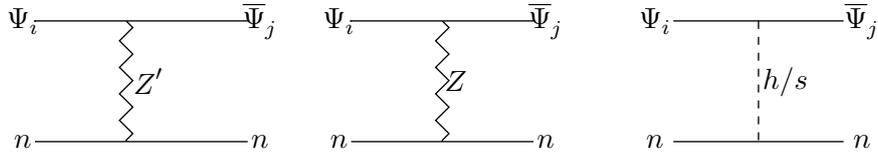
\begin{figure}[H]
	\centering
   \begin{tikzpicture}[line width=0.5 pt, scale=0.8]
	\draw[solid] (-3.5,1.0)--(-2.0,1.0);
        \draw[solid] (-3.5,-1.0)--(-2.0,-1.0);
        \draw[snake](-2.0,1.0)--(-2.0,-1.0);
        \draw[solid] (-2.0,1.0)--(0.0,1.0);
        \draw[solid] (-2.0,-1.0)--(0.0,-1.0);
        \node at (-3.7,1.0) {${\Psi_i}$};
        \node at (-3.7,-1.0) {$n$};
        \node [right] at (-2.05,0.0) {$Z'$};
        \node at (0.2,1.0) {$\overline{\Psi}_j$};
        \node at (0.2,-1.0) {$n$};

         \draw[solid] (1.7,1.0)--(3.2,1.0);
        \draw[solid] (1.7,-1.0)--(3.2,-1.0);
        \draw[snake](3.2,1.0)--(3.2,-1.0);
        \draw[solid] (3.2,1.0)--(4.7,1.0);
        \draw[solid] (3.2,-1.0)--(4.7,-1.0);
        \node at (1.5,1.0) {${\Psi_i}$};
        \node at (1.5,-1.0) {$n$};
        \node [right] at (3.07,0.0) {$Z$};
        \node at (4.9,1.0) {$\overline{\Psi}_j$};
        \node at (4.9,-1.0) {$n$};

         \draw[solid] (7.0,1.0)--(8.4,1.0);
        \draw[solid] (7.0,-1.0)--(8.4,-1.0);
        \draw[dashed](8.4,1.0)--(8.4,-1.0);
        \draw[solid](8.4,1.0)--(9.8,1.0);
        \draw[solid] (8.4,-1.0)--(9.8,-1.0);
        \node at (6.7,1.0) {${\Psi_i}$};
        \node at (6.7,-1.0) {$n$};
        \node [right] at (8.3,0.0) {$h/s$};
        \node at (10.1,1.0) {$\overline{\Psi}_j$};
        \node at (10.1,-1.0) {$n$};
     \end{tikzpicture}
\caption{Scattering of DM with SM particles via $Z$, $Z'$ gauge Bosons and via $h$, $s$ scalars. }
\label{fig:fig13}
 \end{figure}

\subsection{Feynman diagrams relevant with relic density}
\label{app:Cb}
\begin{figure}[H]
\centering
    \begin{tikzpicture}[line width=0.5 pt, scale=1.35]
	\draw[solid] (-3.5,1.0)--(-2.0,1.0);
        \draw[solid] (-3.5,-0.5)--(-2.0,-0.5);
        \draw[solid](-2.0,1.0)--(-2.0,-0.5);
        \draw[snake] (-2.0,1.0)--(0.0,1.0);
        \draw[snake] (-2.0,-0.5)--(0.0,-0.5);
        \node at (-3.7,1.0) {${\Psi_i}$};
        \node at (-3.7,-0.5) {$\overline{\Psi}_j$};
        \node [right] at (-2.05,0.25) {${\Psi^+}$};
        \node at (0.21,1.0) {$W^+$};
        \node at (0.21,-0.5) {$W^-$};

         \draw[solid] (1.7,1.0)--(3.2,1.0);
        \draw[solid] (1.7,-0.5)--(3.2,-0.5);
        \draw[solid](3.2,1.0)--(3.2,-0.5);
        \draw[snake] (3.2,1.0)--(4.7,1.0);
        \draw[snake] (3.2,-0.5)--(4.7,-0.5);
        \node at (1.5,1.0) {${\Psi_i}$};
        \node at (1.5,-0.5) {$\overline{\Psi}_j$};
        \node [right] at (3.07,0.25) {${\Psi_{i,j}}$};
        \node at (4.9,1.0) {$Z$};
        \node at (4.9,-0.5) {$Z$};

     \end{tikzpicture}

    \begin{tikzpicture}[line width=0.5 pt, scale=1.35]
	\draw[solid] (-3.5,1.0)--(-2.0,1.0);
        \draw[solid] (-3.5,-0.5)--(-2.0,-0.5);
        \draw[solid](-2.0,1.0)--(-2.0,-0.5);
        \draw[dashed] (-2.0,1.0)--(0.0,1.0);
        \draw[dashed] (-2.0,-0.5)--(0.0,-0.5);
        \node at (-3.7,1.0) {${\Psi_i}$};
        \node at (-3.7,-0.5) {$\overline{\Psi}_j$};
        \node [right] at (-2.05,0.25) {${\Psi_{i,j}}$};
        \node at (0.2,1.0) {$h/s$};
        \node at (0.2,-0.5) {$h/s$};

         \draw[solid] (1.7,1.0)--(3.2,1.0);
        \draw[solid] (1.7,-0.5)--(3.2,-0.5);
        \draw[solid](3.2,1.0)--(3.2,-0.5);
        \draw[snake] (3.2,1.0)--(4.7,1.0);
        \draw[dashed] (3.2,-0.5)--(4.7,-0.5);
        \node at (1.5,1.0) {${\Psi_i}$};
        \node at (1.5,-0.5) {$\overline{\Psi}_j$};
        \node [right] at (3.07,0.25) {${\Psi_{i,j}}$};
        \node at (4.9,1.0) {$Z$};
        \node at (4.9,-0.5) {$h/s$};

     \end{tikzpicture}

\vspace{0.1in}
\centering
    \begin{tikzpicture}[line width=0.5 pt, scale=0.85]
          \draw[solid] (-3.0,1.0)--(-1.5,0.0);
        \draw[solid] (-3.0,-1.0)--(-1.5,0.0);
         \draw[snake] (-1.5,0.0)--(0.8,0.0);
         \draw[snake] (0.8,0.0)--(2.5,1.0);
         \draw[snake] (0.8,0.0)--(2.5,-1.0);
         \node at (-3.3,1.0) {${\Psi_i}$};
         \node at (-3.3,-1.0) {$\overline{\Psi}_j$};
         \node [above] at (0.0,0.05) {$Z$};
         \node at (2.9,1.0) {$W^+$};
        \node at (2.9,-1.0) {$W^-$};
         \draw[solid] (5.0,1.0)--(6.5,0.0);
        \draw[solid] (5.0,-1.0)--(6.5,0.0);
         \draw[dashed] (6.5,0.0)--(8.5,0.0);
         \draw[snake] (8.5,0.0)--(10.1,1.0);
         \draw[snake] (8.5,0.0)--(10.1,-1.0);
         \node at (4.7,1.0) {${\Psi_i}$};
         \node at (4.7,-1.0) {$\overline{\Psi}_j$};
         \node [above] at (7.4,0.05) {$h/s$};
         \node at (10.4,1.0) {$W^+$};
        \node at (10.4,-1.0) {$W^-$};
     \end{tikzpicture}
\vspace{0.1in}
\centering
    \begin{tikzpicture}[line width=0.5 pt, scale=0.85]
          \draw[solid] (-3.0,1.0)--(-1.5,0.0);
        \draw[solid] (-3.0,-1.0)--(-1.5,0.0);
         \draw[dashed] (-1.5,0.0)--(0.8,0.0);
        \draw[solid] (0.8,0.0)--(2.5,1.0);
         \draw[solid] (0.8,0.0)--(2.5,-1.0);
         \node at (-3.3,1.0) {${\Psi_i}$};
         \node at (-3.3,-1.0) {$\overline{\Psi}_j$};
         \node [above] at (0.0,0.05) {$h/s$};
        \node at (2.7,1.0) {$q$};
        \node at (2.7,-1.0) {$\overline{q}$};
         \draw[solid] (5.0,1.0)--(6.5,0.0);
        \draw[solid] (5.0,-1.0)--(6.5,0.0);
         \draw[dashed] (6.5,0.0)--(8.5,0.0);
         \draw[dashed] (8.5,0.0)--(10.4,1.0);
         \draw[dashed] (8.5,0.0)--(10.4,-1.0);
         \node at (4.7,1.0) {${\Psi_i}$};
         \node at (4.7,-1.0) {$\overline{\Psi}_j$};
         \node [above] at (7.4,0.05) {$h/s$};
         \node at (10.7,1.0) {$h/s$};
        \node at (10.7,-1.0) {$h/s$};
     \end{tikzpicture}

\begin{tikzpicture}[line width=0.5 pt, scale=0.85]
          \draw[solid] (-3.0,1.0)--(-1.5,0.0);
        \draw[solid] (-3.0,-1.0)--(-1.5,0.0);
         \draw[dashed] (-1.5,0.0)--(1.0,0.0);
         \draw[snake] (1.0,0.0)--(3.0,1.0);
         \draw[snake] (1.0,0.0)--(3.0,-1.0);
         \node at (-3.3,1.0) {${\Psi_i}$};
         \node at (-3.3,-1.0) {$\overline{\Psi}_j$};
         \node [above] at (0.0,0.05) {$h/s$};
         \node at (3.2,1.0) {$Z$};
        \node at (3.2,-1.0) {$Z$};
         \draw[solid] (5.0,1.0)--(6.5,0.0);
        \draw[solid] (5.0,-1.0)--(6.5,0.0);
         \draw[snake] (6.5,0.0)--(8.5,0.0);
         \draw[snake] (8.5,0.0)--(10.4,1.0);
         \draw[dashed] (8.5,0.0)--(10.4,-1.0);
         \node at (4.7,1.0) {${\Psi_i}$};
         \node at (4.7,-1.0) {$\overline{\Psi}_j$};
         \node [above] at (7.4,0.05) {$Z$};
         \node at (10.7,1.0) {$Z$};
        \node at (10.7,-1.0) {$h/s$};
     \end{tikzpicture}

\caption{Annihilation ($i=j$) and co-annihilation($i\neq j$) of DM states to Higgs, exotic scalar, quarks and gauge bosons final states are mediated via the Standard Model Higgs, exotic scalar, and vector bosons. The diagrams in the top two rows represent t-channel processes mediated via exotic fermion states. Here, ($i,j = 1,2$).}
\label{fig:fig2}
 \end{figure}
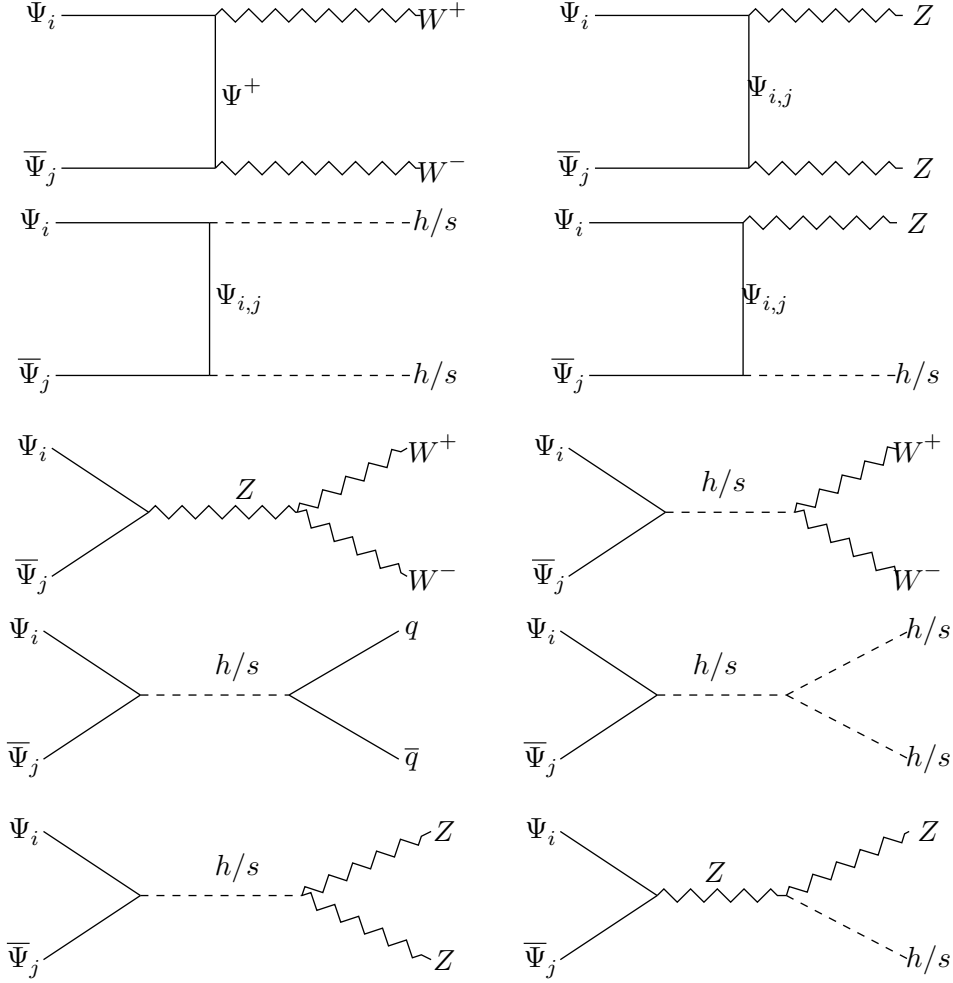

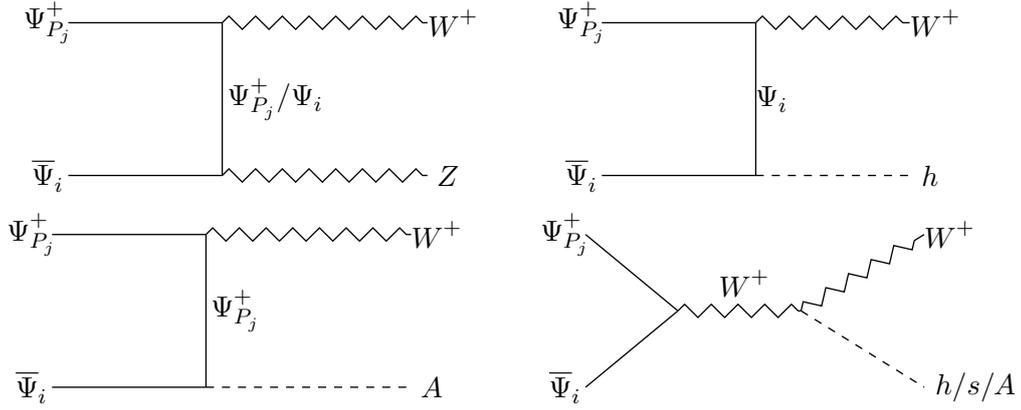
\begin{figure}[H]
\centering
\begin{tikzpicture}[line width=0.5 pt, scale=1.35]
	\draw[solid] (-3.5,1.0)--(-2.0,1.0);
        \draw[solid] (-3.5,-0.5)--(-2.0,-0.5);
        \draw[solid](-2.0,1.0)--(-2.0,-0.5);
        \draw[snake] (-2.0,1.0)--(0.0,1.0);
        \draw[snake] (-2.0,-0.5)--(0.0,-0.5);
        \node at (-3.7,1.0) {${\Psi_{P_j}^+}$};
        \node at (-3.7,-0.5) {$\overline{\Psi}_i$};
        \node [right] at (-2.05,0.25) {${\Psi_{P_j}^+}/{\Psi_{i}}$};
        \node at (0.25,1.0) {$W^+$};
        \node at (0.2,-0.5) {$Z$};

        \draw[solid] (1.7,1.0)--(3.2,1.0);
        \draw[solid] (1.7,-0.5)--(3.2,-0.5);
        \draw[solid](3.2,1.0)--(3.2,-0.5);
        \draw[snake] (3.2,1.0)--(4.7,1.0);
        \draw[dashed] (3.2,-0.5)--(4.7,-0.5);
        \node at (1.5,1.0) {${\Psi_{P_j}^+}$};
        \node at (1.5,-0.5) {$\overline{\Psi}_i$};
        \node [right] at (3.1,0.25) {${\Psi_{i}}$};
        \node at (4.95,1.0) {$W^+$};
        \node at (4.9,-0.5) {$h$};

     \end{tikzpicture}

    \begin{tikzpicture}[line width=0.5 pt, scale=1.35]
	\hspace{0.1in}
	\draw[solid] (-3.5,1.0)--(-2.0,1.0);
        \draw[solid] (-3.5,-0.5)--(-2.0,-0.5);
        \draw[solid](-2.0,1.0)--(-2.0,-0.5);
        \draw[snake] (-2.0,1.0)--(0.0,1.0);
        \draw[dashed] (-2.0,-0.5)--(0.0,-0.5);
        \node at (-3.7,1.0) {${\Psi_{P_j}^+}$};
        \node at (-3.7,-0.5) {$\overline{\Psi}_i$};
        \node [right] at (-2.05,0.25) {${\Psi_{P_j}^+}$};
        \node at (0.25,1.0) {$W^+$};
        \node at (0.2,-0.5) {$A$};

         \draw[solid] (1.7,1.0)--(2.6,0.25);
        \draw[solid] (1.7,-0.5)--(2.6,0.25);
        \draw[snake](2.6,0.25)--(3.8,0.25);
        \draw[snake] (3.8,0.25)--(5.0,1.0);
        \draw[dashed] (3.8,0.25)--(5.0,-0.5);
        \node at (1.5,1.0) {${\Psi_{P_j}^+}$};
        \node at (1.5,-0.5) {$\overline{\Psi}_i$};
        \node [right] at (2.9,0.52) {$W^+$};
        \node at (5.25,1.0) {$W^+$};
        \node at (5.5,-0.5) {$h/s/A$};

     \end{tikzpicture}
\caption{Co-annihilation processes of $\Psi_i$ ($i=1,2$) DM states with the charged exotic fermions $\Psi_{P_j}^+$ ($j=1,2$) to Higgs, exotic scalar, and SM gauge boson final states for various t-channel and s-channel processes. }
\label{fig:fig3}
\end{figure}
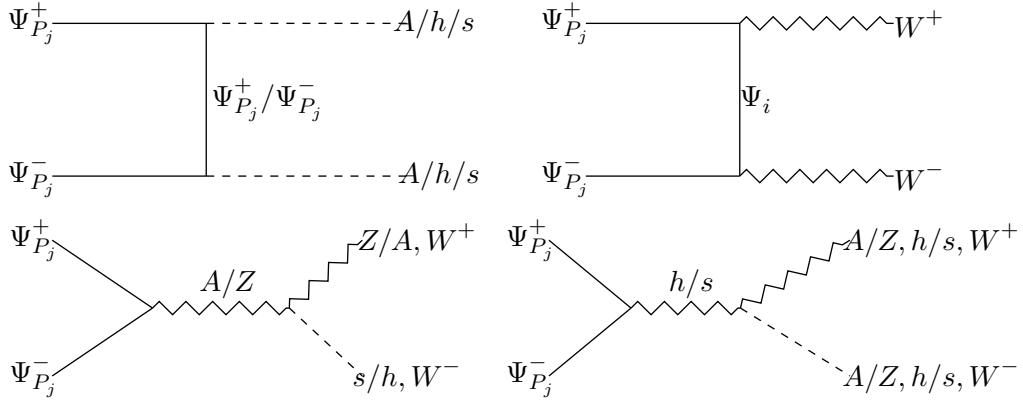
\begin{figure}[H]
\centering
\begin{tikzpicture}[line width=0.5 pt, scale=1.35]
	\draw[solid] (-3.5,1.0)--(-2.0,1.0);
        \draw[solid] (-3.5,-0.5)--(-2.0,-0.5);
        \draw[solid](-2.0,1.0)--(-2.0,-0.5);
        \draw[dashed] (-2.0,1.0)--(0.0,1.0);
        \draw[dashed] (-2.0,-0.5)--(0.0,-0.5);
        \node at (-3.7,1.0) {$\Psi_{P_j}^+$};
        \node at (-3.7,-0.5) {$\Psi^-_{P_j}$};
        \node [right] at (-2.05,0.25) {${\Psi_{P_j}^+}/{\Psi^-_{P_j}}$};
        \node at (0.22,1.0) {$A/h/s$};
        \node at (0.27,-0.5) {$A/h/s$};

        \draw[solid] (1.7,1.0)--(3.2,1.0);
        \draw[solid] (1.7,-0.5)--(3.2,-0.5);
        \draw[solid](3.2,1.0)--(3.2,-0.5);
        \draw[snake] (3.2,1.0)--(4.7,1.0);
        \draw[snake] (3.2,-0.5)--(4.7,-0.5);
        \node at (1.5,1.0) {$\Psi_{P_j}^+$};
        \node at (1.5,-0.5) {$\Psi^-_{P_j}$};
        \node [right] at (3.1,0.25) {${\Psi_{i}}$};
        \node at (4.95,1.0) {$W^+$};
        \node at (4.95,-0.5) {$W^-$};

     \end{tikzpicture}
     \hspace{-1.2in}
    \begin{tikzpicture}[line width=0.5 pt, scale=1.2]
        \draw[solid] (-2.9,1.0)--(-1.8,0.25);
        \draw[solid] (-2.9,-0.5)--(-1.8,0.25);
        \draw[snake](-1.8,0.25)--(-0.3,0.25);
        \draw[snake] (-0.3,0.25)--(0.5,1.0);
        \draw[dashed] (-0.3,0.25)--(0.5,-0.5);
        \node at (-3.10,1.0) {${\Psi_{P_j}^+}$};
        \node at (-3.10,-0.5){${\Psi_{P_j}^-}$};
        \node [right] at (-1.4,0.52) {$A/Z$};
        \node at (1.1,1.0) {$Z/A,W^+$};
        \node at (0.99,-0.5) {$s/h,W^-$};
       \hspace{0.4in}
         \draw[solid] (1.7,1.0)--(2.6,0.25);
        \draw[solid] (1.7,-0.5)--(2.6,0.25);
        \draw[snake](2.6,0.25)--(3.8,0.25);
        \draw[snake] (3.8,0.25)--(5.0,1.0);
        \draw[dashed] (3.8,0.25)--(5.0,-0.5);
        \node at (1.5,1.0) {${\Psi_{P_j}^+}$};
        \node at (1.5,-0.5){${\Psi_{P_j}^-}$};
        \node [right] at (2.9,0.52) {$h/s$};
        \node at (5.89,1.0) {$A/Z,h/s,W^+$};
        \node at (5.89,-0.5) {$A/Z,h/s,W^-$};

     \end{tikzpicture}
\caption{Co-annihilation processes of the charged exotic fermions $\Psi_{P_j}^\pm$ $(j=1,2)$ to Higgs, exotic scalar, and SM gauge bosons final states for various t-channel and s-channel processes.}
\label{fig:fig3w}
\end{figure}

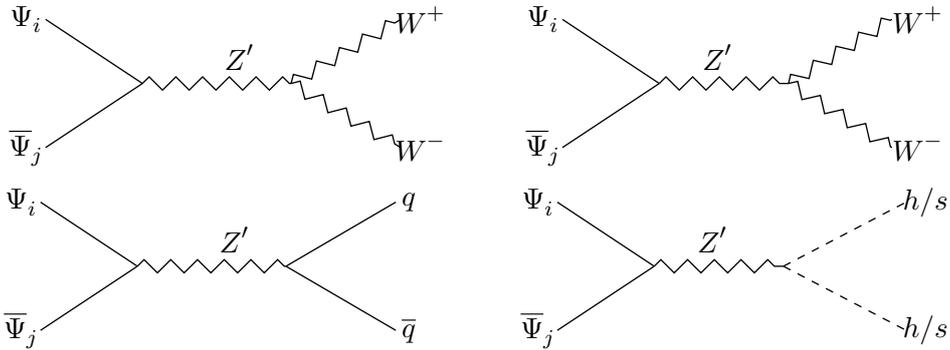
\begin{figure}[H]
\centering
    \begin{tikzpicture}[line width=0.5 pt, scale=0.85]
          \draw[solid] (-3.0,1.0)--(-1.5,0.0);
        \draw[solid] (-3.0,-1.0)--(-1.5,0.0);
         \draw[snake] (-1.5,0.0)--(0.8,0.0);
         \draw[snake] (0.8,0.0)--(2.5,1.0);
         \draw[snake] (0.8,0.0)--(2.5,-1.0);
         \node at (-3.3,1.0) {${\Psi_i}$};
         \node at (-3.3,-1.0) {$\overline{\Psi}_j$};
         \node [above] at (0.0,0.05) {$Z'$};
         \node at (2.8,1.0) {$W^+$};
        \node at (2.8,-1.0) {$W^-$};
         \draw[solid] (5.0,1.0)--(6.5,0.0);
        \draw[solid] (5.0,-1.0)--(6.5,0.0);
         \draw[snake] (6.5,0.0)--(8.5,0.0);
         \draw[snake] (8.5,0.0)--(10.1,1.0);
         \draw[snake] (8.5,0.0)--(10.1,-1.0);
         \node at (4.7,1.0) {${\Psi_i}$};
         \node at (4.7,-1.0) {$\overline{\Psi}_j$};
         \node [above] at (7.4,0.05) {$Z'$};
         \node at (10.5,1.0) {$W^+$};
        \node at (10.5,-1.0) {$W^-$};
     \end{tikzpicture}
\centering
    \begin{tikzpicture}[line width=0.5 pt, scale=0.85]
          \draw[solid] (-3.0,1.0)--(-1.5,0.0);
        \draw[solid] (-3.0,-1.0)--(-1.5,0.0);
         \draw[snake] (-1.5,0.0)--(0.8,0.0);
        \draw[solid] (0.8,0.0)--(2.5,1.0);
         \draw[solid] (0.8,0.0)--(2.5,-1.0);
         \node at (-3.3,1.0) {${\Psi_i}$};
         \node at (-3.3,-1.0) {$\overline{\Psi}_j$};
         \node [above] at (0.0,0.05) {$Z'$};
        \node at (2.7,1.0) {$q$};
        \node at (2.7,-1.0) {$\overline{q}$};
         \draw[solid] (5.0,1.0)--(6.5,0.0);
        \draw[solid] (5.0,-1.0)--(6.5,0.0);
         \draw[snake] (6.5,0.0)--(8.5,0.0);
         \draw[dashed] (8.5,0.0)--(10.4,1.0);
         \draw[dashed] (8.5,0.0)--(10.4,-1.0);
         \node at (4.7,1.0) {${\Psi_i}$};
         \node at (4.7,-1.0) {$\overline{\Psi}_j$};
         \node [above] at (7.4,0.05) {$Z'$};
         \node at (10.7,1.0) {$h/s$};
        \node at (10.7,-1.0) {$h/s$};
     \end{tikzpicture}
\caption{Annihilation processes~($i=j$) of fermionic states to Higgs and gauge bosons final states. }
\label{fig:fig4}
 \end{figure}
\begin{figure}[H]
\centering
    \begin{tikzpicture}[line width=0.5 pt, scale=0.85]
        \draw[solid] (-3.5,1.0)--(-2.0,1.0);
        \draw[solid] (-3.5,-1.0)--(-2.0,-1.0);
        \draw[solid](-2.0,1.0)--(-2.0,-1.0);
        \draw[dashed] (-2.0,1.0)--(0.0,1.0);
        \draw[snake] (-2.0,-1.0)--(0.0,-1.0);
        \node at (-3.7,1.0) {${\Psi_i}$};
        \node at (-3.7,-1.0) {$\overline{\Psi}_j$};
        \node [right] at (-2.05,0.0) {${\Psi_{i,j}}$};
        \node at (0.2,1.0) {$h/s$};
        \node at (0.2,-1.0) {$Z'$};
         \draw[solid] (1.7,1.0)--(3.2,1.0);
        \draw[solid] (1.7,-1.0)--(3.2,-1.0);
        \draw[solid](3.2,1.0)--(3.2,-1.0);
        \draw[snake] (3.2,1.0)--(4.7,1.0);
        \draw[snake] (3.2,-1.0)--(4.7,-1.0);
        \node at (1.5,1.0) {${\Psi_i}$};
        \node at (1.5,-1.0) {$\overline{\Psi}_j$};
        \node [right] at (3.07,0.0) {${\Psi_{i,j}}$};
        \node at (4.9,1.0) {$Z'$};
        \node at (4.9,-1.0) {$Z'$};

         \draw[solid] (7.0,1.0)--(8.4,1.0);
        \draw[solid] (7.0,-1.0)--(8.4,-1.0);
        \draw[solid](8.4,1.0)--(8.4,-1.0);
        \draw[snake](8.4,1.0)--(9.8,1.0);
        \draw[dashed] (8.4,-1.0)--(9.8,-1.0);
        \node at (6.7,1.0) {${\Psi_i}$};
        \node at (6.7,-1.0) {$\overline{\Psi}_j$};
        \node [right] at (8.17,0.0) {${\Psi_{i,j}}$};
        \node at (10.1,1.0) {$Z'$};
        \node at (10.2,-1.0) {$h/s$};
     \end{tikzpicture}
\centering
    \begin{tikzpicture}[line width=0.5 pt, scale=0.85]
          \draw[solid] (-3.0,1.0)--(-1.5,0.0);
        \draw[solid] (-3.0,-1.0)--(-1.5,0.0);
         \draw[snake] (-1.5,0.0)--(1.0,0.0);
         \draw[solid] (1.0,0.0)--(3.0,1.0);
         \draw[solid] (1.0,0.0)--(3.0,-1.0);
         \node at (-3.3,1.0) {${\Psi_i}$};
         \node at (-3.3,-1.0) {$\overline{\Psi}_j$};
         \node [above] at (0.0,0.05) {$Z'$};
         \node at (3.3,1.0) {$q$};
        \node at (3.3,-1.0) {$\overline{q}$};
         \draw[solid] (5.0,1.0)--(6.5,0.0);
        \draw[solid] (5.0,-1.0)--(6.5,0.0);
         \draw[snake] (6.5,0.0)--(8.5,0.0);
         \draw[snake] (8.5,0.0)--(10.4,1.0);
         \draw[snake] (8.5,0.0)--(10.4,-1.0);
         \node at (4.7,1.0) {${\Psi_i}$};
         \node at (4.7,-1.0) {$\overline{\Psi}_j$};
         \node [above] at (7.4,0.05) {$Z'$};
         \node at (10.8,1.0) {$W^+$};
        \node at (10.8,-1.0) {$W^-$};
     \end{tikzpicture}
\centering
    \begin{tikzpicture}[line width=0.5 pt, scale=0.85]
          \draw[solid] (-3.0,1.0)--(-1.5,0.0);
        \draw[solid] (-3.0,-1.0)--(-1.5,0.0);
         \draw[dashed] (-1.5,0.0)--(1.0,0.0);
         \draw[snake] (1.0,0.0)--(3.0,1.0);
         \draw[snake] (1.0,0.0)--(3.0,-1.0);
         \node at (-3.3,1.0) {${\Psi_i}$};
         \node at (-3.3,-1.0) {$\overline{\Psi}_j$};
         \node [above] at (0.0,0.05) {$h/s$};
         \node at (3.3,1.0) {$Z'$};
        \node at (3.3,-1.0) {$Z'$};
         \draw[solid] (5.0,1.0)--(6.5,0.0);
        \draw[solid] (5.0,-1.0)--(6.5,0.0);
         \draw[snake] (6.5,0.0)--(8.5,0.0);
         \draw[snake] (8.5,0.0)--(10.4,1.0);
         \draw[dashed] (8.5,0.0)--(10.4,-1.0);
         \node at (4.7,1.0) {${\Psi_i}$};
         \node at (4.7,-1.0) {$\overline{\Psi}_j$};
         \node [above] at (7.4,0.05) {$Z'$};
         \node at (10.7,1.0) {$Z'$};
        \node at (10.7,-1.0) {$h/s$};
     \end{tikzpicture}
\caption{Dominant co-annihilation processes ($i \ne j$) to Higgs, exotic scalar, quarks, and other SM and exotic gauge boson final states. Processes in the first row are t-channel, mediated by neutral exotic fermion states, with final states containing at least one exotic gauge boson. Processes in the last two rows are s-channel, containing at least one exotic gauge boson either in the final state or as a mediator.}
\label{fig:fig5}
 \end{figure}
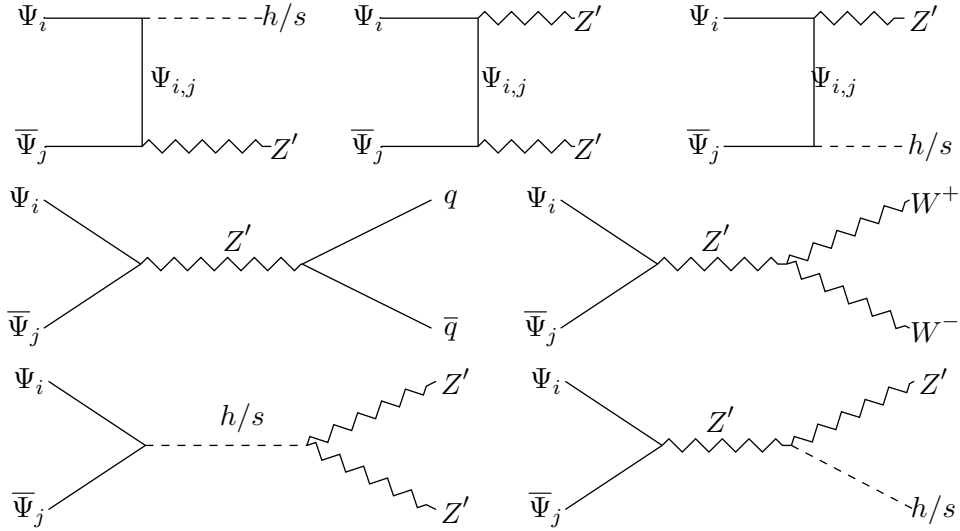

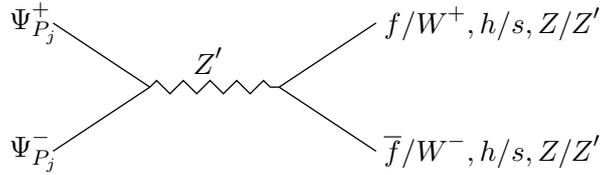
\begin{figure}[H]
\centering
    \begin{tikzpicture}[line width=0.5 pt, scale=0.85]
         \draw[solid] (5.0,1.0)--(6.5,0.0);
        \draw[solid] (5.0,-1.0)--(6.5,0.0);
         \draw[snake] (6.5,0.0)--(8.5,0.0);
         \draw[solid] (8.5,0.0)--(10.0,1.0);
         \draw[solid] (8.5,0.0)--(10.0,-1.0);
         \node at (4.7,1.0) {$\Psi_{P_j}^{+}$};
         \node at (4.7,-1.0) {$\Psi_{P_j}^{-}$};
         \node [above] at (7.4,0.05) {$Z'$};
         \node at (11.8,1.0) {$f/W^+,h/s,Z/Z'$};
        \node at (11.8,-1.0) {$\overline{f}/W^-,h/s,Z/Z'$};
     \end{tikzpicture}
\caption{Co-annihilation processes of charged fermions~($\Psi_{P_j}^{\pm}$) with $(j=1,2)$ to SM particles in final states mediated via exotic gauge boson $Z'$. }
\label{fig:fig6}
 \end{figure}
\bibliographystyle{JHEP}
\bibliography{dm}
\end{document}